%% file: main.tex
\documentclass[sigplan]{acmart}

\renewcommand\footnotetextcopyrightpermission[1]{} % removes footnote with conference information in first column
\usepackage{epsfig,endnotes,url,color}
\usepackage{array}
\usepackage[english]{babel}
\usepackage{xspace}
\usepackage{verbatim}
\usepackage{amsmath}
\usepackage{caption}
\usepackage{fancyhdr}
\usepackage[caption=false,font=footnotesize]{subfig}
\usepackage{listings}
\usepackage{multirow}
\usepackage{booktabs}
\usepackage{tikz}
\usepackage{pgfplots}
\usepackage{balance}
\usepackage{pgfplotstable}
\pgfplotsset{compat=1.7}

\usetikzlibrary{shapes.geometric, arrows, fit, positioning, shadows,
   backgrounds, pgfplots.groupplots}

\extrafloats{100}
\usepackage{pgfplots}

\input{mac}

\newcommand{\myfontsize}{\fontsize{8}{9}\selectfont}
\newcommand{\sys}{\textsc{Sieve}\xspace}
\newcommand{\kshape}{$k$-Shape\xspace}

\lstdefinelanguage{javascript}{
  keywords={break, case, catch, continue, debugger, default, delete, do, else,
finally, for, function, if, in, instanceof, new, return, switch, this, throw,
try, typeof, var, void, while, with},
  basicstyle=\footnotesize,
  morecomment=[l]{//},
  morecomment=[s]{/*}{*/},
  morestring=[b]',
  morestring=[b]",
  sensitive=true
}

\begin{document}

\copyrightyear{2017}
\acmYear{2017}
\setcopyright{acmlicensed}
\acmConference{Middleware '17}{December 11--15, 2017}{Las Vegas, NV, USA}\acmPrice{15.00}\acmDOI{10.1145/3135974.3135977}
\acmISBN{978-1-4503-4720-4/17/12}

\author{J{\"o}rg Thalheim$^{1}$, Antonio Rodrigues$^{2}$, Istemi Ekin Akkus$^3$, Pramod Bhatotia$^1$,\\Ruichuan Chen$^3$,   Bimal Viswanath$^{4}$, Lei Jiao$^{5}$, Christof Fetzer$^6$}
\affiliation{$^{1}$University of Edinburgh, $^{2}$Carnegie Mellon Univ., $^{3}$NOKIA Bell Labs, $^{4}$University of Chicago, $^{5}$University of Oregon, $^{6}$TU Dresden}

%\author{J{\"o}rg Thalheim}
%\affiliation{\institution{University of Edinburgh}}
%%\email{joerg@thalheim.io}
%
%\author{Antonio Rodrigues}
%\affiliation{\institution{Carnegie Mellon University}}
%%\email{adamiaonr@cmu.edu}
%
%\author{Ekin Akkus}
%\affiliation{\institution{NOKIA Bell Labs}}
%%\email{}
%
%\author{Pramod Bhatotia}
%\affiliation{\institution{University of Edinburgh}}
%%\email{pramod.bhatotia@ed.ac.uk}
%
%\author{Ruichuan Chen}
%\affiliation{\institution{NOKIA Bell Lab}}
%%\email{}
%
%\author{Bimal Viswanath}
%\affiliation{\institution{University of Chicago}}
%%\email{viswanath@cs.uchicago.edu}
%
%\author{Lei Jiao}
%\affiliation{\institution{University of Oregon}}
%%\email{}
%
%\author{Christof Fetzer}
%\affiliation{\institution{TU Dresden}}
%%\email{}

\renewcommand{\shortauthors}{Thalheim et al.}

% \author{ J{\"o}rg Thalheim$^{\dag+}$, Antonio Rodrigues$^{\diamond+}$,
% Istemi Ekin Akkus$^\ddag$, Pramod Bhatotia$^\#$,\\Ruichuan Chen$^\ddag$,   Bimal Viswanath$^{\star+}$, Lei Jiao$^{\bullet+}$, Christof Fetzer$^\dag$\\%,
% \small $^\dag$TU Dresden, $^\diamond$Carnegie Mellon University, $^\ddag$NOKIA Bell Labs, \\ 
% \small $^\#$The University of Edinburgh,  $^\star$University of California Santa Barbara, $^\bullet$University of Oregon \\
% \small $^+$Work done mostly while at NOKIA Bell Labs.}

\renewcommand{\shortauthors}{Thalheim et al.}

\title{Sieve: Actionable Insights from \\Monitored Metrics in Microservices} % in Microservices}

\input{abstract}

%   <ccs2012>
%   <concept>
%   <concept_id>10010520.10010575</concept_id>
%   <concept_desc>Computer systems organization~Dependable and fault-tolerant systems and networks</concept_desc>
%   </concept>
%   </ccs2012>

% \ccsdesc[500]{Computer systems organization~Dependable and fault-tolerant
% systems and networks}

%\keywords{Microservices, Time series analysis\\{\footnotesize $^*$Authors did part of the work at NOKIA Bell Labs.}}

\maketitle{}

\input{introduction}

\input{overview}

\input{design}
\input{case-study} % implementation of case-studies
\input{implementation}
\input{evaluation}

\input{related-work}

\input{lessons}
\input{conclusion}

\balance

\bibliographystyle{abbrv}
\bibliography{main}

\end{document}

%% file: mac.tex
\newcommand{\myparagraph}[1]{\smallskip \noindent{\bf {#1}.}}
\newcommand{\out}[1] {}

\newcounter{codeLineCntr}

\reversemarginpar
\newif\ifnotes
\notestrue

\newcommand{\punt}[1]{}

\renewcommand{\eqref}[1]{Equation~(\ref{eq:#1})}

\newcommand{\proc}[1]{\ifmmode\mbox{\textsc{#1}}\else\textsc{#1}\fi}

  \newcommand{\func}[1]{\ifmmode\mathrm{#1}\else\textrm{#1}fi} %
\newcounter{remark}[section]

%% file: abstract.tex
\begin{abstract}

Major cloud computing operators provide powerful monitoring tools to understand the current (and prior) state of the distributed systems deployed in their infrastructure. While such tools provide a detailed monitoring mechanism at scale, they also pose a significant challenge for the application developers/operators to transform the  {\em huge space of monitored metrics} into {\em useful insights}. These insights are essential to build effective management tools for improving the efficiency, resiliency, and dependability of distributed systems.  

% for improving the efficiency, resiliency, and dependability of distributed systems.  

%. These insights are essential to build effective management tools for a wide-range of workflows to  improve the efficiency, resiliency, and dependability of distributed systems.  

This paper reports on our experience with building and deploying \sys{}---a platform to derive actionable insights from monitored metrics in distributed systems.  \sys{} builds on two core components: a metrics reduction framework, and a metrics dependency extractor. More specifically, \sys{} first reduces the dimensionality of metrics by automatically filtering out unimportant metrics by observing their signal over time. % using a time-series clustering algorithm.  %To achieve this goal, \sys{} uses a time-series clustering algorithm to group highly related metrics of an application into groups and select a representative metric from each group to reduce the overall amount of metrics. 
Afterwards, \sys{} infers metrics dependencies between distributed components of the system using a predictive-causality model by testing for Granger Causality.  

We implemented \sys{} as a generic platform and deployed it for two microservices-based distributed systems: OpenStack and ShareLatex. 
Our experience shows that (1) \sys{} can reduce the number of metrics by at least an order of magnitude ($10-100\times$), while preserving the statistical equivalence to the total number of monitored metrics; (2)  \sys can dramatically improve existing monitoring infrastructures by reducing the associated overheads over the entire system stack (CPU---80\%, storage---90\%, and network---50\%); 
(3) 
% \sys can easily adapt to the variations in system workloads (and also components) to produce valid consistent information for the evaluated distributed system; (4) 
Lastly, \sys{} can be effective to support a wide-range of workflows in distributed systems---we showcase two such workflows: Orchestration of autoscaling, and Root Cause Analysis (RCA). This technical report is an extended version of our conference publication~\cite{sieve-middleware}.

%The system is designed and deployed in the context of micro-services research at NOKIA Bell Labs.
%\pramod{add some key results from the evaluation}

%Most distributed systems are constantly monitored to understand their current
%and prior state, and this monitoring is a crucial part of any system deployment. 
%
%In this respect, many distributed systems applications are designed
%following the microservices architecture. These applications are split up
%into smaller services that can be deployed individually, and communicate with
%each other over well-defined network based APIs. In the current setting, the
%number of services and metrics for such systems can grow beyond the
%understanding of a single developer or operator. 
%
%In this paper we present
%\sys{}---a metric reduction framework for microservices. \sys{} decreases the
%dimensionality of metrics needed to considered. \sys automatically filters
%unimportant metrics by observing their signal over time. \sys{} uses a novel
%time-series clustering algorithm called K-Shape to group highly related metrics of a
%service into groups and select a representative metric from each group to
%reduce the overall amount of metrics. \sys{} infers dependencies between service
%components using a predictive-causality model by testing for Granger Causality.
%We show that \sys's generic approach is useful to support two case-studies: auto-scaling and root-cause analysis in micro-services. 

\end{abstract}

%% file: introduction.tex
\section{Introduction}

Most distributed systems are constantly monitored to understand their
current (and prior) states.
The
main purpose of monitoring is to gain \textit{actionable insights} that would
enable a developer/operator to take appropriate actions to better manage the deployed
system. Such insights are commonly used to manage the health and
resource requirements as well as to investigate and recover from
failures (root cause identification). For these reasons, 
monitoring is a crucial part of any distributed system deployment.

All major cloud computing operators provide a monitoring infrastructure for application
developers (e.g., Amazon CloudWatch~\cite{cloudwatch}, Azure Monitor~\cite{azure-monitor}, Google StackDriver~\cite{stackdriver}). 
These platforms provide infrastructure to monitor a large number (hundreds or thousands) of various application-specific and system-level metrics associated with a cloud application. 
Although such systems feature scalable measurement and
storage frameworks to conduct monitoring at scale, they leave the task of 
transforming the monitored metrics into usable knowledge to the developers. 
Unfortunately, this transformation becomes difficult 
with the increasing size and complexity of the application.

In this paper, we share our experience on: 
{\em How can we derive actionable insights from the monitored metrics in distributed systems?} 
In particular, given a large number of monitored metrics across different components (or processes) in a distributed system, we want to design a platform that can derive actionable insights from the monitored metrics. This platform could be used to support a wide-range of use cases to improve the 
efficiency, resiliency, and reliability of distributed systems.

In this work, we focus on microservices-based distributed systems because they have become the de-facto way to design and deploy 
modern day large-scale web applications~\cite{martinfowler-microservices}. 
The microservices architecture is an ideal candidate for our study for two reasons:
%The microservices architecture is an ideal candidate to investigate the complexities in distributed systems for two reasons:
First, microservices-based applications have a large number of distributed components (hundreds to thousands \cite{uber-architecture, netflix-microscope}) with complex communication patterns, each component usually exporting several metrics for the purposes of debugging, performance diagnosis, and application management. 
Second, microservices-based applications are developed at a rapid pace:
% following the DevOps paradigm: 
new features are being continuously integrated and deployed. Every new update may fix some existing issues, 
introduce new features, but can also introduce a new bug. With this rapid update schedule, keeping track of the changes in the application as a whole with effects propagating to other
components becomes critical for reliability, efficiency, and management purposes.

The state-of-the-art management infrastructures either rely on ad hoc techniques or custom application-specific tools. 
For instance, prior work in this space has mostly focused on analyzing message-level traces
(instead of monitored metrics) to generate a causal model of the application to debug performance issues~\cite{mysterymachine, blackboxes}. 
Alternatively, developers 
% of microservices 
usually create and use custom tools to address the complexity of understanding the application
as a whole. For example, Netflix developed several application-specific tools for such purposes~\cite{netflix-microscope, netflix-vector}
by instrumenting the entire application.
These approaches require either complicated instrumentation or
sophisticated techniques to infer happens-before relationships (for the causal
model) by analyzing message trace timestamps, making them inapplicable for broader use.
% It would be ideal to derive actionable
% insights from already monitored data without intruding the application
% code. 
%Fortunately, many application deployments already have a monitoring framework that collects such data.
%In our work, we simply rely on
%\textit{any available} monitored information to derive actionable insights and
%it is well known that most app deployments come with a monitoring framework.

% Consequently, understanding the dependencies between the components
% and utilizing these dependencies with the exported metrics becomes difficult. 
% To address the issue, the developers of microservices usually create ad hoc tools following the
% DevOps paradigm. For example, the application programmers at Netflix developed several application-specific tools for such purposes~\cite{netflix-microscope, netflix-vector}. These tools, however, require the
% application under investigation to be instrumented, 
% so that the communication
% pattern between components can be established by following requests coming into
% the application. This kind of instrumentation may not be possible for service
% operators, who may not have access to the application code. Furthermore, to take
% advantage of such a tool for following requests, the developers would have to
% agree and follow the convention, which can be a limiting factor for modularity.

This paper presents our experience with designing and
building \sys, a system that can 
utilize an existing monitoring infrastructure (i.e., without changing the monitored information) 
to infer actionable insights for application
management. 
% By leveraging the advances made in the data mining community,
\sys takes a data-driven approach to enable better management of microservices-based
applications. At its core, \sys{} is composed of two key modules: (1) a
metric reduction engine that reduces the dimensionality of the metric space by
filtering out metrics that carry redundant information, (2) a metric dependency
extractor that builds a causal model of the application by inferring causal
relationships between metrics associated with different components. 

\if 0
The metric reduction is not a
straight-forward task and requires a time-series clustering algorithm that is
resistant to noise and other types of distortion. \sys employs a recent algorithm, \kshape \cite{kshape},
for this task. The dependency extractor needs to be agnostic to 
application and metric semantics. \sys relies on a statistical causality inference technique
called \textit{Granger Causality} \cite{grangercausality}.
\fi
%\sys's
%dependency extractor is agnostic to application and metric semantics and relies
%on a statistical causality inference technique called \textit{Granger
%Causality} \cite{grangercausality}. 

Module (1)\label{1} enables \sys{} to identify ``relevant'' metrics for a given
application management task. For instance, it might be sufficient to monitor
only a few metrics associated with error states of the application instead of
the entire set when monitoring the health of the application. %, reducing the monitoring overhead in the network and storage. 
It is important to also note that reducing the metric space has implications for 
deployment costs:
frameworks like Amazon CloudWatch use a per-metric charging model, and not identifying relevant
metrics can significantly drive up the cost related to monitoring the application. 

Module (2)\label{2} is crucial for inferring actionable
insights because it is able to automatically infer complex application
dependencies. In a rapidly updating application, the ability to observe such
complex dependencies and how they may change is important for keeping one's
understanding of the application as a whole up-to-date. Such up-to-date
information can be helpful for developers to quickly react to any problem that
may arise during deployment.
We implemented \sys as a generic platform, and deployed it with two microservices-based distributed systems: ShareLatex \cite{sharelatex} and  OpenStack \cite{openstack-kolla}.
Our experience shows that (1) \sys{} can reduce the number of monitored metrics by an order of magnitude ($10-100\times$), while preserving the statistical equivalence to the total number of monitored metrics. In this way, the developers/operators can focus on the important metrics that actually matter. (2)  \sys can dramatically improve the efficiency of existing  metrics monitoring infrastructures by reducing the associated overheads over the entire system stack (CPU---80\%, storage---90\%, and network---50\%).  This is especially important for systems deployed in a cloud infrastructure, where the monitoring infrastructures (e.g. AWS CloudWatch) charge customers for monitoring resources. And finally, (3) \sys can be employed for supporting a wide-range of workflows. We showcase two such case-studies:  In the first case study, we use ShareLatex \cite{sharelatex} and 
show how \sys can help developers orchestrate autoscaling of microservices-based applications. 
In the second case study, we use OpenStack \cite{openstack-kolla} and 
show how developers can take advantage
of \sys's ability to infer complex dependencies across various components in microservices for Root Cause Analysis (RCA).  \sys{}'s source code with the full experimentation setup is publicly available: \href{https://sieve-microservices.github.io/}{https://sieve-microservices.github.io/}.

%% file: overview.tex
\section{Overview}%Motivation, Goals and Assumptions}
In this section, we first present some background on microservices-based applications and our motivation
to focus on them. Afterwards, we present our goals,  design overview, and its possible use cases.

 \subsection{Background and Motivation}
 %In this work, we focus on microservices-based distributed systems. 
 Microservices-based applications consist of loosely-coupled distributed components (or processes) that communicate via well-defined interfaces. 
 %Typical microservices-based
 %applications are composed of hundreds of components \cite{uber-architecture,
 %netflix-microscope}. 
 Designing and building applications in this way increases
 modularity, so that developers can work on different components and maintain
 them independently. These advantages 
make the microservices architecture the de facto design choice for large-scale web 
applications~\cite{martinfowler-microservices}.

%While increasing modularity, such an approach to developing software can also increase the application complexity: As the number of components increases, the interconnections between components also increases. 
While increasing modularity, such an approach to developing software can also increase the application complexity: As the number of components increases, the interconnections between components also increases. Furthermore, each component usually exports several metrics for the purposes of debugging, performance diagnosis, and
application management. Therefore, understanding the dependencies between the components
 and utilizing these dependencies with the exported metrics becomes a challenging task. As a result, understanding how the application performs as a whole becomes increasingly
difficult.

 \input{tables/tab-motivation}

\input{figureTex/fig-design}

Typical microservices-based
 applications are composed of hundreds of components \cite{uber-architecture,
 netflix-microscope}. Table~\ref{table:tab-motivation} shows real-world microservices-based applications that have tens of thousands of metrics and hundreds of components. We experimented with two such applications, ShareLatex~\cite{sharelatex} and OpenStack~\cite{openstack}, each having several thousands of metrics and order of tens of components. The metrics in these applications come from all layers of the application like hardware counters, resource usage, business metrics or application-specific metrics.

%Table~\ref{table:tab-motivation} shows real-world microservice deployments that have tens of thousands of metrics and hundreds of components. We experimented with two microservices, ShareLatex~\cite{sharelatex} and OpenStack~\cite{openstack}, each having order of tens of components and several thousands of metrics. These metrics come from all layers of the service like hardware counters, resource usage, business metrics or application specific metrics.

To address this data overload issue, developers of microservices-based applications usually create ad hoc tools.
%following the DevOps paradigm. 
For example, application programmers at Netflix developed several application-specific tools for such purposes~\cite{netflix-microscope, netflix-vector}. These tools, however, require the
 application under investigation to be instrumented, so that the communication
 pattern between components can be established by following requests coming into
 the application. This kind of instrumentation requires coordination among developers of different
 components, which can be a limiting factor for modularity.
 
% This kind of instrumentation may not be possible for service
% operators, who may not have access to the application code. Furthermore, to take
% advantage of such a tool for following requests, the developers would have to
% agree and follow the convention, which can be a limiting factor for modularity.
 
 Major cloud computing operators also provide monitoring tools for recording all metric
 data from all components. For example,  Amazon CloudWatch~\cite{cloudwatch}, Azure Monitor~\cite{azure-monitor}, and Google StackDriver~\cite{stackdriver}. 
 These monitoring tools aid in visualizing and processing metric data in real-time (i.e., for
 performance observation) or after an issue with the application (i.e., for debugging). These tools, however, either use a few system metrics that
 are hand-picked by developers based on experience, or simply record all metric data for all the components.
 %because they do not know how to filter redundant and less useful metrics
 %beforehand. 
 %These monitoring tools try to visualize and process the data
 %after an issue with the application (i.e., for debugging) or real-time (i.e., for
 %performance observation). These tools, however, either rely on few system metrics that
 %are hand-picked via experience, or recording all metric data for all components,
 %because they do not know how to filter redundant and less useful metrics
 %beforehand. 
 %Although such processing may help the developers learn more about
 %their application components, it is not clear whether the service needs to
 %record and collect all metric data from all components all the time the
 %application is running.

Relying on past experience may not always be effective due
to the increasing complexity of a microservices-based application. On the other
hand, recording all metric data can create significant monitoring overhead in the network
and storage, or in the case of running the application in a cloud infrastructure
(e.g., AWS), it can incur costs due to the provider charging the customers
(e.g., CloudWatch). For these reasons, it is important to understand the
dependencies between the components of a microservice-based application.
Ideally, this process should not be intrusive to the application. Finally, it
should help the developers to identify and minimize the critical components and
metrics to monitor.

\subsection{Design Goals}
\label{subsec:goals}
%\myparagraph{Design goals} 
While designing \sys, we set the following goals.

\begin{itemize}
\item {\bf Generic:} Many tools for distributed systems have specific goals, including
performance debugging, root cause analysis and orchestration. Most of the time, these tools
are custom-built for the application in consideration and target a certain goal.
%Besides
%targeting certain goals, these tools are also custom-built for the application in mind most of the time.
Our goal is to design a generic platform that can be used for a wide-range of workflows.

\item {\bf  Automatic:} The sheer number of metrics prohibits manual inspection. On the other
hand, designing a generic system to help developers in many use cases might require
manually adjusting some parameters for each use case. Our tool should be as automated as possible
while reducing the number of metrics and extracting their relationships. However, we leave
the utilization of our platform's output to the developers, who may have different goals.

\item {\bf Efficient:} Our platform's operation should be as efficient as possible. Minimizing analysis time
becomes important when considering distributed systems, such as microservices-based applications.

\end{itemize}

\myparagraph{Assumptions} While developing \sys, we made two assumptions.
% First, we assume that the developers have 
% access to the underlying infrastructure, so that they can install and run software on the servers running the
% components of the applications. In microservices-based applications following a DevOps model, 
% this assumption is reasonable because the developers and the operators of the applications are the same or belong to the same organization.
\begin{itemize}
\item We assume that the developers can supply a workload generator for the application under
investigation. This assumption should be reasonable and less demanding for developers than fully instrumenting each component and/or figuring out relationships across all metrics.
\item It is possible for specific use cases to have additional assumptions. 
For example, a root cause
analysis scenario would require knowing a faulty and non-faulty (correct) version of the application.

\end{itemize}

% Second, we assume that the infrastructure and the application already expose several metrics, so that 
% we do not need to instrument the application for this information. Fortunately, many
% components in microservices-based applications already export several metrics that include general system (e.g., CPU, memory)
% as well as application-specific metrics (e.g., request latency, number of error messages).

%<<<<<<< HEAD
%\item It is possible for specific use cases to have additional assumptions. 
%For example, a root cause
%analysis scenario would require knowing a faulty and non-faulty version of the application.
%=======
%>>>>>>> 262e212d4ac25e09cd988aefe5f96d082107ed45
%
%\end{itemize}
%\myparagraph{Assumptions} We make the following assumptions (1) access to the underlying
%hosting infrastructure. (2) annotations about different metrics and types (3)
%application metrics are already exposed by the developers and doesn't require
%further changes (4) 

\subsection{\sys Overview}
%\bimal{Do read the two paragraphs below. made some changes. previous text is commented out.}
%The basic idea of \sys is to systematically analyze and filter collected
%metrics and to identify dependencies between components. 
The underlying intuition behind \sys is two-fold: Firstly, in the
metric dimension, some metrics of a component may behave with similar patterns
as other metrics of that component. Secondly, in the component dimension, there are dependencies between components. As a result, monitoring
all metrics of all components at runtime may be unnecessary and inefficient (as components are not independent).

%Furthermore, the number of metrics exposed and the number of components present
%in the application makes this monitoring more complex, creating problems and
%conflicts during operation (e.g., when making scaling decisions) or development
%(e.g., trying to debug an issue). 

%The basic idea of \sys is to reduce the complexity of understanding the
%component dependencies by systematically analyzing and filtering collected
%metrics as well as identifying the critical components and their relationships
%with regards the other components. The underlying intuition is twofold: Firstly, in the
%metric dimension, some metrics of a component may behave with similar patterns
%as other metrics of that component. Secondly, in the component dimension, not
%all components may be as critical as some components. As a result, monitoring
%all metrics of all components at runtime may be unnecessary and inefficient.
%Furthermore, the number of metrics exposed and the number of components present
%in the application makes this monitoring more complex, creating problems and
%conflicts during operation (e.g., when making scaling decisions) or development
%(e.g., trying to debug an issue). 

In this paper, we present \sys to reduce
this complexity by systematically analyzing the application to filter collected
metrics and to build a dependency graph across components. To showcase the generality of this dependency graph and its benefits, we then utilize \sys to orchestrate autoscaling of the ShareLatex~\cite{sharelatex} application---an online collaboration tool,
and to perform Root Cause Analysis (RCA) in OpenStack~\cite{openstack-kolla}---a cloud management software (\S \ref{sec:case-study}).

%In this paper, we present \sys to reduce
%this complexity by systematically analyzing the application to filter collected
%metrics and to identify critical components, so that a dependency graph across
%the components can be created. To showcase the generality of this dependency graph
%and its benefits, we then utilize \sys in orchestration of autoscaling using ShareLatex~\cite{sharelatex}, an
%online collaboration tool,
%and performing root cause analysis (RCA) in OpenStack~\cite{openstack-kolla}, a cloud management software
%(\S \ref{sec:case-study}).

At a high level, \sys's design follows three steps as shown in Figure~\ref{fig:design-figure}.

%\paragraph{Step #1: Load the application, and record communications and
%metrics.} 
\myparagraph{Step \#1: Load the application} \sys uses an application-specific load generator to 
stress the application under investigation. 
This load generator can be provided by the application developers.
%similar to unit tests. 
For example, OpenStack already uses a load generator named Rally
\cite{openstack-rally}. During the load, \sys records the communications among
components to obtain a {\em call graph}. This recording does not require any
modifications to the application code. In addition, \sys records all exposed
{\em metrics} by all components. Note that this recording only happens during the
creation of the call graph and not during runtime.

\myparagraph{Step \#2: Reduce metrics}
After collecting
the metrics, \sys analyzes each component and organizes its metrics into fewer
groups via clustering, so that similar-behaving metrics are clustered together.
After clustering, \sys picks a representative metric from each cluster. These
representative metrics as well as their clusters in a sense characterize each
component.

\myparagraph{Step \#3: Identify dependencies} 
In this step,
\sys explores the possibilities of one component's representative metrics
affecting another component's metrics using a pairwise comparison method: each
representative metric of one component is compared with each representative
metric of another component. \sys uses the call graph obtained in Step 1 to choose the components to be compared (i.e., components directly communicating) and the
representative metrics determined in Step 2. As a result, the search space is
significantly reduced compared to the na\"ive approach of comparing all
components with every other component using all metrics.

% can be removed
%\iea{I don't think we can remove the following text, because Granger causality gives us 'relationships'
%(i.e., predictive causality or predictive happens-before relationship) and this text describes the connection
%between such a relationship and a dependency.}
If \sys determines that there is a relationship between a metric of one
component and another metric of another component, a dependency edge between
these components is created using the corresponding metrics. The direction of
the edge depends on which component is affecting the other. 
%The incoming number
%of edges to a component determine the criticality of the component: the fewer
%the incoming edges, the more critical the component.

%\myparagraph{Potential use cases} We envision \sys can be useful to the developers or operators of distributed systems to build a wide-range of management tools to improve the efficiency, reliability, and resiliency of distributed systems. While we showcase the useful  Each of these cases might require some tweaks and specific knowledge about the application. Nevertheless, we think that the output of \sys can be a starting point. In Section \ref{sec:case-study}, we showcase two of these use cases with two different applications.

\input{use_cases}

%% file: tables/tab-motivation.tex
\begin{table}[t]
\caption{Number of metrics exposed by microservices.}
%\small
\myfontsize
\centering
\newcommand{\tabitem}{~~~~~\llap{\textbullet}~}
\begin{tabular}{@{}lcccc@{}}
\toprule
Microservices                    & Number of metrics \\ \midrule
Netflix~\cite{netflix-atlas}                     & $\sim$ 2,000,000 \\
Quantcast~\cite{quantcast}                   & $\sim$ 2,000,000 \\
Uber~\cite{uber-monitoring}                  & $\sim$ 500,000,000 \\
\hline
ShareLatex~\cite{sharelatex}                  & 889  \\
OpenStack~\cite{openstack-apis, openstack-telemetry}                & 17,608  \\
\bottomrule
\end{tabular}

\label{table:tab-motivation}
\end{table}

%% file: figureTex/fig-design.tex
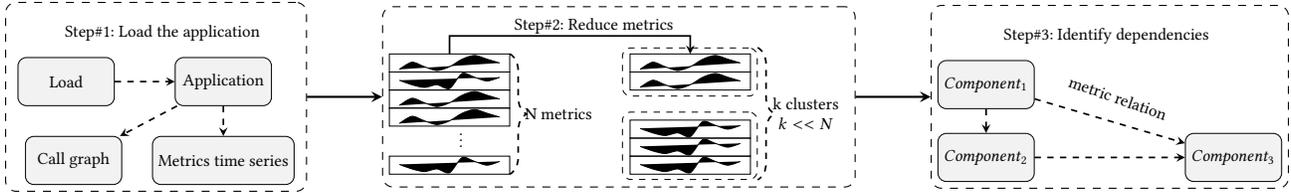
\begin{figure*}[t]
\centering
\input{figures/design-figure}
\caption{High level steps of \sys.}
\label{fig:design-figure}
%\vspace*{-0.4cm}
\end{figure*}

%% file: figures/design-figure.tex
\pgfplotstableread[row sep=crcr]{
1   0.14 \\
2   0.33 \\
3   0.21 \\
4   0.11 \\
5   0.21 \\
6   0.41 \\
7   0.51 \\
8   0.41 \\
9   0.31 \\
}\tablea

\pgfplotstableread[row sep=crcr]{
1   0.33 \\
2   0.21 \\
3   0.14 \\
4   0.21 \\
5   0.11 \\
6   0.51 \\
7   0.41 \\
8   0.31 \\
9   0.41 \\
}\tableb

\tikzstyle{box} = [rectangle,
  rounded corners,
  minimum width=5cm,
  minimum height=1cm,
  text centered,
  draw=black,
  dashed
]
\tikzstyle{process} = [rectangle,
  minimum width=2cm,
  minimum height=1cm,
  text centered,
  draw=black,
  fill=gray!10
]
\tikzstyle{arrow} = [thick,->,>=stealth]

\newsavebox{\myplot}
\sbox{\myplot}{%
\begin{tikzpicture}[every node/.style={scale=0.8},font=\large]
    \pgfplotsset{mystyle/.style=
       scale only axis,
       xtick={1,2,3},
       ticks=none,
       axis line style={solid},
       axis x line=box,
       axis y line=box,
       width=2cm,
       height=0.3cm,
       every axis plot/.style={
         ybar,
         fill=gray!10,
         smooth,
         color=black,
         mark=none,
         solid,
       }
    }
    \begin{groupplot}[mystyle,
         group style={
           group name=left plots,
           group size=1 by 5,
           vertical sep=0pt
         }]
      \nextgroupplot
      \addplot[mystyle] table {\tablea};
      \nextgroupplot
      \addplot[mystyle] table {\tableb};
      \nextgroupplot
      \addplot[mystyle] table {\tablea};
      \nextgroupplot
      \addplot[mystyle] table {\tablea} node(g1) {};

      \nextgroupplot[yshift=-0.5cm]
      \addplot[mystyle] table {\tableb} node(g2) {};

    \end{groupplot}
    \draw[thick,decorate,decoration={brace,amplitude=8pt}]
      (left plots c1r1.north east) --
      (left plots c1r5.south east)
      node[midway, right,xshift=0.2cm]{N metrics};

    \node[rotate=90,yshift=0.8cm] at ($(g1)!.5!(g2)$) {\ldots};

    \begin{groupplot}[mystyle,
        group style={
          group name=right plots,
          group size=1 by 5,
          vertical sep=0pt
        }]
      \nextgroupplot[anchor=north west, at={($(left plots c1r1.north east) + (2cm,0)$)}]
      \addplot[mystyle] table {\tablea} node(groupa1) {};
      \nextgroupplot
      \addplot[mystyle] table {\tablea} node(groupa2) {};

      \nextgroupplot[yshift=-0.5cm]
      \addplot[mystyle] table {\tableb} node(groupb1) {};
      \nextgroupplot[]
      \addplot[mystyle] table {\tableb} node(groupb2) {};
      \nextgroupplot[]
      \addplot[mystyle] table {\tableb} node(groupb3) {};
    \end{groupplot}
    \node [box,
      fit=(groupa1) (groupa2),
      minimum width=2.8cm,
      minimum height=1cm,
      xshift=-1.05cm,
    ] (groupa) {};
    \node [box,
      fit=(groupb1) (groupb2) (groupb3),
      minimum width=2.8cm,
      minimum height=1.4cm,
      xshift=-1.05cm,
      yshift=-0.05cm,
    ] (groupb) {};

    \draw[thick,decorate,decoration={brace,amplitude=8pt}]
      (groupa.north east) --
      (groupb.south east)
      node[midway, right,xshift=0.2cm,align=center]{
        k clusters\\
        $k << N$
       };
    \draw[arrow,solid] (left plots c1r1.north) -- +(0,.3) -| (right plots
    c1r1.north) node[above,sloped,yshift=0.3cm,xshift=-2cm]{Step\#2: Reduce metrics};
    %\node[above=0.3cm of clusterarrow] 
    %  at ($(clusterarrow.west)!.3!(clusterarrow.east)$) {};
  \end{tikzpicture}
}

\begin{tikzpicture}[scale=0.2,every node/.style={scale=0.8},font=\large]

\begin{scope}[on background layer]
\node [box] (load) {
  \begin{tikzpicture}
     \node [solid, process] (app) {Application};
     \node [solid, process, left=of app] (req) {Load};
     \node [solid, process, below=0.5cm of app] (metrics) {Metrics time series};
     \node [solid, process, left=0.5cm of metrics] (callgraph) {Call graph};
     \node [solid, above=0cm of req,xshift=2cm] (title_1) {Step\#1: Load the application};
     \draw[arrow] (req) -- (app);
     \draw[arrow] (app) -- (metrics);
     \draw[arrow] (app) -- (callgraph);
  \end{tikzpicture}
};
\end{scope}

\node [box, right=of load] (reduce) {
\usebox{\myplot}
};

\node [box, right=of reduce] (identify) {
  \begin{tikzpicture}
    \node (title3) {Step\#3: Identify dependencies};
    \node[ellipse, solid, process, below=0cm of title3, xshift=-2.5cm] (c1) {$Component_1$};
    \node[ellipse, solid, process, below=0.4cm of c1] (c2) {$Component_2$};
    \node[ellipse, solid, process, right=2.5cm of c2] (c3) {$Component_3$};
    \draw[arrow] (c1) -- (c2);
    \draw[arrow] (c1) -- (c3) node [midway,above,sloped,rotate=-3]{metric relation};
    \draw[arrow] (c2) -- (c3);
  \end{tikzpicture}
};

\draw[arrow] (load) -- (reduce);
\draw[arrow] (reduce) -- (identify);
\end{tikzpicture}

%% file: use_cases.tex
\subsection{Potential Use Cases}

We envision \sys can be useful to the developers or operators of distributed systems to build a wide-range of management tools to improve the efficiency, reliability, and resiliency of distributed systems. Each of these cases might require some tweaks and specific knowledge about the application. Nevertheless, we think that the output of \sys can be a starting point. In Section \ref{sec:case-study}, we showcase two of these use cases with two different applications.

\myparagraph{Orchestration and autoscaling of components} The pairwise investigation of representative metrics of components produces the dependencies across components. 
%These dependencies also cover the time lag when one component starts seeing the changes after the first component as well as the scaling factor that can be used to add or remove instances of components as needed. 
By definition, the dependency graph shows the order of bottlenecked components. 
As a result of this graph, developers 
can have a better understanding of which components need to be scaled out first, meaning that the
number of monitored components can be reduced.
%that need to be monitored. one only requires to monitor the roots of the dependency graph to make scaling decisions, reducing the number of monitored components at runtime. 
Furthermore, the dependencies show the metrics that are affected, meaning that one only needs to monitor a limited set of metrics rather than every metric exported by these components. In combination, these reductions are reflected in the monitoring overhead. Also, scaling decisions are based on fewer components and metrics, so that potentially conflicting scaling decisions can be avoided.

\myparagraph{Root cause analysis} It is possible that updates can introduce bugs and problems into the application. Changing dependency graphs (i.e., after updates) may indicate that the problem got introduced during a particular update that caused the dependency graph to change. Identifying such changes will be useful in debugging the application by pointing to the root cause of the problem.

\myparagraph{Performance diagnosis} Similar to the root cause analysis scenario, the resource-usage profile of a component may change after an update. Even if the interfaces between the components may stay the same, the update may address a previously unknown bug, introduce a new feature and/or implement the same functionality in a more efficient and/or a different way. Such changes can have effects on other parts of the application, and the complexity of the application may render these effects not easily foreseeable. The dependency graph can be utilized to understand the overall effect on the application the update can have.
% and can help with ``what if?" analyses.

\myparagraph{Placement decisions} When scaling decisions are made, the resource-usage profiles of components become important, because components with similar resource-usage profiles may create contention (e.g., two components may be CPU-bound). As a result, placing a new instance of a component on a computer where other instances of the same component or instances of another component with a similar resource-usage profile run may not yield the additional benefit of the extra instance. The dependency graph lists which metrics are affected in a component under load. This information can be used to extract resource-usage profiles of components to avoid such contention during scale-up events.

%% file: design.tex
\section{Design}
%
%\sys aims to reduce the complexity of understanding
%the component dependencies by systematically analyzing and filtering collected
%metrics as well as identifying the critical components and their relationships
%regarding the other components. 

In this section, we detail the three steps of \sys.

%\pramod{Remove: The underlying intuition is twofold: In the
%metric dimension, some metrics of a component may behave with similar patterns
%as other metrics of that component. Similarly, in the component dimension, not
%all components may be as critical as others.}

\subsection{Load the Application}
\label{sec:design:profiling}
For our systematic analysis, we first run the application under various
load conditions. This loading serves two purposes: First, the load exposes a number
of metrics from the application as well as the infrastructure it runs on. These metrics are
then used to identify potential relationships across components.
Second, the load also enables us to obtain a
call graph, so that we can identify the components that communicate with each other. The
call graph is later used to reduce the amount of computation required to identify
the inter-component relationships ($\S$\ref{sec:design:relationships}).
The load test is intended to be run in an offline step and not in production.

\myparagraph{Obtaining metrics}
During the load of the application, we record metrics as time series.
%These time series will then be
%used to identify potential relationships across components. Our goal is to
%understand how metrics develop under load and which metrics follow the same trend over
%time.
There are two types of metrics that we can leverage for our analysis:
%to understand the relationships across components. 
First, there are system metrics that are obtained from the underlying
operating system. These metrics report the resource usage of a microservice component,
and are usually related to the hardware resources on a host. Examples include
usages in CPU, memory, network and disk I/O.

Second, there are application-level metrics. Application developers often add
application-specific metrics (e.g., number of active users, response time of a request
in a component). Commonly-used components (e.g., databases, load balancers) and
certain language runtimes (e.g., Java) may provide statistics about specific operations
(e.g., query times, request counts, duration of garbage collection).

\myparagraph{Obtaining the call graph}
\label{ssec:callgraph}
Generally speaking, applications using a microservices architecture communicate
via well-defined interfaces similar to remote procedure calls.
We model these communications between the components as a directed graph,
where the vertices represent the microservice components and the edges
point from the caller to the callee providing the service.

By knowing which components communicate directly, we can reduce the number of
component pairs we need to check to see whether they have a relation (see Section
\ref{sec:design:relationships}). Although it is possible to manually track this
information for smaller-sized applications, this process becomes quickly
difficult and error-prone with increasing number of components.

There are several ways to understand which microservice components are
communicating with each other. One can instrument the application, so that each
request can be traced from the point it enters the application to the point
where the response is returned to the user. Dapper~\cite{dapper} from Google and
Atlas~\cite{netflix-atlas, netflix-microscope} from Netflix rely on instrumenting their RPC middleware to trace
requests.

Another method to obtain communicating components is to monitor network
traffic between hosts running those components using a tool like \emph{tcpdump}.
After obtaining the traffic, one can map the exchanged packets to
the components via their source/destination addresses. 
This method can
produce communicating component pairs by parsing all network packets,
adding significant computational overhead and increasing the analysis time.
Furthermore, it is possible that many
microservice components are deployed onto the same host (e.g., using containers), 
making the packet parsing difficult due to network address translation on the host machine.

One can also observe system calls related to network operations via APIs such as
\emph{ptrace()} \cite{ptrace}. However, this approach adds a lot of context
switches between the tracer and component under observation. 
%the added overhead can be significant, leading to metrics being skewed.
% can skew
% the metrics observed, leading to
% significant overhead, and increases the time of the analysis.

\sys employs \emph{sysdig} to obtain the communicating pairs.
\emph{sysdig}\cite{sysdig} is a recent project providing a new method to observe
system calls in a more efficient way. Utilizing a kernel module, \emph{sysdig}
provides system calls as an event stream to a user application. The event stream
also contains information about the monitored processes, so that network
calls can be mapped to microservice components, even if they are running
in containers. Furthermore, it enables extraction of the communication peer via
user-defined filters. Employing \emph{sysdig}, we avoid the shortcomings of the
above approaches: 1) We do not need to instrument the application, which makes our
system more generally applicable, 2) We add little overhead to obtain the call graph of an
application for our analysis (see Section
\ref{subsubsec:monitoring}).

\subsection{Reduce Metrics}
\label{sec:design:metric_reduction}
The primary goal of exporting metrics is to understand
the performance of applications, orchestrating them and debugging
them. While the metrics exported by the application developers or commonly-used
microservice components may be useful for these purposes, it is often the case that the developers
have little idea regarding which ones are going to be most useful. Developers
from different backgrounds may have different opinions: a developer specializing
in network communications may deem
network I/O as the most important metric to consider, whereas a developer with a
background on algorithms may find CPU usage more valuable. As a result of these
varying opinions, often times many metrics are exported.

While it may look like there is no harm in exporting as much information as possible
about the application, it can create problems.
Manually investigating the obtained metrics from a large number of components
becomes increasingly difficult
with the increasing number of metrics and components \cite{curse-of-dimensionality}.
This complexity reflects on
the decisions that are needed to control and maintain the application. In addition,
the overhead associated with the collection and storage of these metrics can quickly
create problems. In fact, Amazon CloudWatch \cite{cloudwatch} charges its customers for the
reporting of the metrics they export. As a result, the more metrics an application has
to export, the bigger the cost the developers would have to bear.

One observation we make is that some metrics strongly correlate with each other
and it might not be necessary to consider all of them when making decisions about
the control of the application. For example, some application metrics might
be strongly correlated with each other due to the redundancy in choosing which metrics
to export by the developers. It is also possible that different subsystems in the same
component report similar information (e.g., overall memory vs. heap
usage of a process). In addition, some system metrics may offer clues
regarding the application's state: increased network I/O may indicate an increase
in the number of requests.

The direct outcome of this observation is that it should be possible to reduce the
dimensionality of the metrics the developers have to consider. As such, the procedure
to enable this reduction should happen with minimal user effort and
%while being able to
%tolerate different amplitudes and shifts in time as well as 
scale with increased numbers
of metrics.

To achieve these requirements, \sys uses a clustering approach named \kshape
\cite{kshape} with a pre-filtering step. While other approaches such as principal
component analysis (PCA) \cite{pca} and random projections \cite{papadimitriou2000}
can also be used for dimensionality reduction, these approaches either produce results
that are not easily interpreted by developers (i.e., PCA) or sacrifice accuracy to achieve
performance and have stability issues producing different results across runs
(i.e., random projections). On the other hand, clustering results can be
visually inspected by developers, who can also use any application-level
knowledge to validate their correctness.
Additionally, clustering can also uncover hidden relationships which might not have
been obvious.

\myparagraph{Filtering unvarying metrics}
Before we use \kshape, we first filter metrics with constant trend or low
variance ($var \leq  0.002$). These metrics cannot provide any new information
regarding the relationships across components, because they are not changing according
to the load applied to the application. Removing these metrics also enables us to improve
the clustering results.

\myparagraph{\kshape clustering}
\kshape is a recent clustering algorithm that scales linearly with the number of metrics.
It uses a novel distance metric called
\emph{shape-based distance} (SBD). SBD is based on a normalized form of
cross correlation (NCC) \cite{kshape}. Cross correlation is calculated
using Fast
Fourier Transformation and normalized using the geometric mean of the
autocorrelation of each individual metric's time series. Given two time series vectors, $\vec{x}$ and $\vec{y}$, SBD will take the position $w$, when sliding $\vec{x}$ over $\vec{y}$, where the normalized cross correlation maximizes.
\begin{align} 
SBD(\vec{x},\vec{y}) &= 1 - max_w \left( NCC_w(\vec{x},\vec{y}) \right)
\end{align}

%SBD will take the $w$ position
%where the normalized cross correlation maximizes.
%\begin{align} NCC(x,y) &= \frac{CC(\vec(x), \vec(y))}{\sqrt{R_0(\vec{x},\vec{x})
%\cdot R_0(\vec{x}, \vec{x})}} \\ SBD(x,y) &= 1 - max_w \left( NCC_w(x,y) \right)
%\end{align}

Because \kshape uses a distance metric based on the shape of the investigated time series,
it can detect
similarities in two time series, even if one lags the other in the time dimension.
This feature
is important to determine relationships across components in microservices-based
applications because a change in one metric in one component may not reflect
on another component's metrics immediately (e.g., due to the network delay of calls between components).

Additionally, \kshape is robust against distortion in amplitude
because data is normalized via z-normalization ($z = \frac{x - \mu}
{\sigma}$) before being processed. This feature is especially important because
different metrics may have different units and thus, may not be directly comparable.

\kshape works by initially assigning time series to clusters randomly. In every iteration, it computes
new cluster centroids according to SBD with the assigned time series. These centroids
are then used to update the assignment for the next iteration until the clusters converge
(i.e., the assignments do not change).

We make three adjustments to employ \kshape in \sys.
First, we preprocess the collected time series to be compatible with \kshape.
\kshape expects the observations to
be equidistantly distributed in the time domain. However, during the load of the
application, timeouts or lost packets can cause gaps between the measurements.

To reconstruct missing data, we use spline interpolation of the third order
(cubic). A spline is defined piecewise by polynomial functions. Compared to
other methods such as averages of previous values or linear interpolation,
spline interpolation provides a higher degree of smoothness. It therefore
introduces less distortion to the characteristics of a time-series~\cite{spline-interpolation}.
%, which has a
%positive effect on our causal-predictive model~\cite{spline-interpolation}.
Additionally, monitoring systems retrieve metrics at different points in time
and need to be discretized to match each other. In order to increase the matching accuracy, we
discretize using $500ms$ instead of the original $2s$ used in the original \kshape
paper \cite{kshape}.

Our second adjustment is to change the initial assignments of metric time series to
clusters. To increase clustering performance and reduce the convergence
overhead, we pre-cluster metrics according to their name similarity (e.g., Jaro
distance \cite{jaro1989}) and use these clusters as the initial assignment
instead of the default random assignment. This adjustment is
reasonable given that many developers use naming conventions when exporting
metrics relating to the same component or resource in question (e.g.,
``cpu\_usage", ``cpu\_usage\_percentile"). The number of iterations to converge
should decrease compared to the random assignment, because similar names
indicate similar metrics. Note that this adjustment is only for performance
reasons; the convergence of the \kshape clustering does not require any
knowledge of the variable names and would not be affected even with a random
initial assignment.

During the clustering process, \kshape requires the number of clusters to be previously determined.
In an application with several components, each of which having various number of metrics, pre-determining
the ideal number of clusters may not be straightforward. Our final adjustment is
to overcome this limitation:
we iteratively vary the number of clusters used by \kshape and pick the number that gives
the best \emph{silhouette value}~\cite{silhouettes}, which is a technique
to determine the quality
of the clusters. 
%It is calculated using the mean of all
%silhouette values over all metrics assigned to
%a particular cluster. 
%For metric $i$, the silhouette value is computed as:
%\begin{align} s(i) = (b(i) - a(i)) / max(a(i), b(i)) \end{align}
%where $a(i)$ is the mean intra-cluster distance and $b(i)$ is the mean nearest-cluster
%distance for metric $i$. 
The silhouette value is $-1$ when the assignment is wrong and $1$ when it is a perfect assignment~\cite{scikit-silhouette-score}. 
We use the SBD as a distance measure in the silhouette computation.

In practice, experimenting with a small number of clusters is sufficient. For our
applications, seven clusters per component was sufficient, where each component
had up to 300 metrics.

\myparagraph{Representative metrics}
After the clustering, each microservice component will have one or more
clusters of metrics.
The number of clusters will most likely be much smaller than the number of metrics
belonging to that component.
Once these clusters are obtained, \sys picks one representative metric from each
cluster.
To pick the representative metric from each cluster, \sys determines the SBD between
each metric and the corresponding centroid of the cluster. The metric with the lowest
metric is chosen as the representative metric for this cluster.

The high-level idea is that the behavior of the cluster will match this representative
metric; otherwise, the rest of the metrics in the cluster would not have been in
the same cluster as this metric.
The set of representative metrics of a component can then be used to describe a
microservice component's behavior. These representative metrics are then used
in conjunction with the call graph obtained in Section \ref{sec:design:profiling}
to identify and understand the relationships across components.

%\iea{I am not sure I follow why this would give us a good representative
%metric. - see wiki, if the shape base distance is a bad metric to get a good
%representive, the whole clustering would be meaningless}
%\joerg {
%On the
%  other hand, if the timing difference is very short, it might be okay to pick a
%  random one.
%}

\subsection{Identify Dependencies}
\label{sec:design:relationships}
To better understand an application, we need to find dependencies across its components.
A na\"ive way of accomplishing this goal would be to compare all components
with each other using all possible metrics. One can clearly see that with the increasing
number of components and metrics, this would not yield an effective
solution.

In the previous section, we described how one can reduce the number of metrics
one has to consider in this pairwise comparison by clustering and obtaining
the representative metrics of each component. Still, comparing all pairs of
components using this reduced set of metrics may be inefficient and redundant
considering the number of components in a typical microservices-based application
(e.g., tens or hundreds).

\sys uses the call graph obtained in Section \ref{sec:design:profiling} to
reduce the number of components that need to be investigated in a pairwise
fashion.  For each component, we do pairwise comparisons using each representative
metric of its clusters with each of its neighbouring components (i.e., callees)
and their representative metrics.

\sys utilizes Granger Causality tests \cite{grangercausality} in this pairwise
comparison. Granger Causality tests are useful in determining whether
a time series can be useful in predicting another time series: In a microservices-based application,
the component interactions closely follow the path a request takes inside the application.
As a result, these interactions can be
predictive of the changes in the metrics of the components in the path. Granger Causality tests offer
a statistical approach in understanding the relationships across these components.
Informally, Granger Causality is defined as follows.
If a metric X is Granger-causing another metric Y, then we can predict Y
better by using the history of both X and Y compared to only using the history
of Y~\cite{casual-definition}.

To utilize Granger Causality tests in \sys, we built two linear models using the ordinary
least-square method \cite{ols}. First, we compare each metric $X_t$ with another metric
$Y_t$. Second, we compare each metric $X_t$ with the time-lagged version of the other
metric $Y_t$: $Y_{t-Lag}$. Covering the cases with a time lag is important because
the load in one component may not be reflected on another component until the second
component receives API calls and starts processing them.

\sys utilizes short delays to build the time-lagged versions of metrics. The reason
is that microservices-based applications typically run in the same data center and their components communicate
over a LAN, where typical round-trip times are in the order of milliseconds.
\sys uses a conservative delay of $500ms$ for unforeseen delays.

To apply the Granger Causality tests and check whether the past values of metric
$X$ can predict the future values of metric $Y$, both models are compared via the F-test \cite{ftest}.
The null hypothesis (i.e., $X$ does not granger-cause $Y$) is rejected if the p-value
is below a critical value.

However, one has to consider various properties of the time series. For example,
the F-test requires the time series to be normally distributed. The load generation
used in Section \ref{sec:design:profiling} can be adjusted to accommodate
this requirement. Also, the F-test might find spurious regressions when
non-stationary time series are
included \cite{superious-regressions}. Non-stationary time series (e.g., monotonically
increasing counters for CPU and network interfaces) can be found
using the Augmented Dickey-Fuller test \cite{adf-test}. For these time series,
the first difference is taken and then used in the Granger Causality tests. Although
longer trends may be lost due to the first difference, accumulating metrics
such as counters do not present interesting relationships for our purposes.

%In order to see if past values of $X$ can predict the future of $Y$
%both models are then compared using F-Test. The NULL hypothesis of this test is
%that X does NOT granger cause Y. We reject the NULL hypothesis, if the p-value
%of the F-Test is below a critical value ($p-value \in \{0\dots1\}$). This test
%requires our time series to be stationary. A time series is
%stationary if it has time independent random distribution. If non-stationary
%time series are included this test might find spurious
%regressions~\cite{superious-regressions}. To check for non-stationarity we used
%the Augmented Dickey–Fuller test. That way we found monotonic increasing
%counters like the CPU time or accumulated packet counter for network interfaces.
%These types of metric have an Order of integration of $I(1)$. To test for
%Granger causality for those metrics, we have to take first difference:
%
%\begin{align}
%  X_n = X_n - X_{n-1}
%\end{align}
%
%Usually long running trends are lost if we take the first difference of time
%series. For the accumulated metrics we have however, we are not interested in
%those relationships. The second requirement for F-Test is that the time series
%are normal distributed. This requirement has been satisfied by adapting the load
%profile of the load generator.

\begin{figure*}[t]
\centering
\includegraphics[scale=.425]{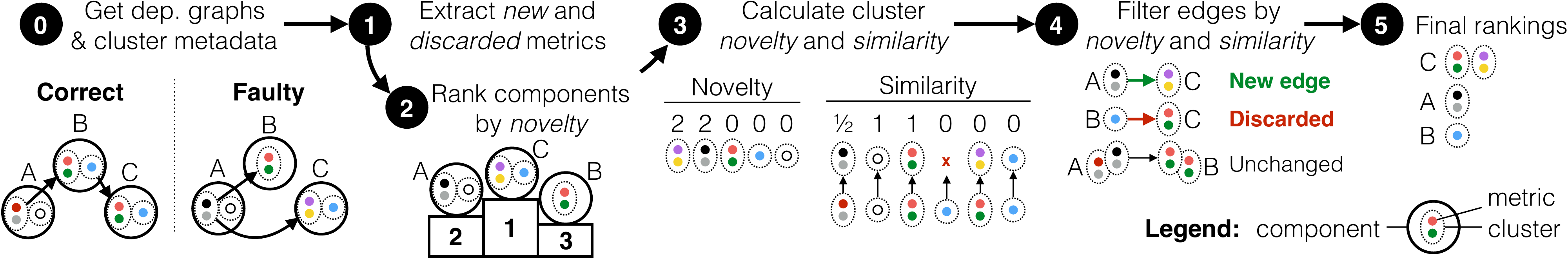}
\caption{\sys's root cause analysis methodology.}
\label{fig:case-study-rootcause-worklow}
\vspace*{-0.4cm}
\end{figure*}

After applying the Granger Causality test to each component's representative
metrics with its neighbouring component's representative metrics, we obtain
a graph. In this graph, we draw an edge between microservice components,
if one metric in one component Granger-causes another metric in a neighbouring
component. This edge represents the dependency between these two components and its direction
is determined by Granger causality.

While Granger Causality tests are useful in determining predictive causality
across microservice components, it has some limitations that we need to consider.
For example, it does not cover instantaneous relationships between two variables.
More importantly, it might reveal spurious relationships, if important variables are
missing in the system: if both $X$ and $Y$ depend on a third variable $Z$ that is
not considered, any relationship found between $X$ and $Y$ may not be useful.
Fortunately, an indicator of such a situation is that both metrics will Granger-cause
each other (i.e., a bidirectional edge in the graph). \sys filters these edges out.

%% file: case-study.tex
\section{Applications}
\label{sec:case-study}

% \label{sec:case-study-autoscaling}

% In addition to \sys's implementation, we built two case-studies on \sys's
% infrastructure to showcase the effectiveness of our approach. Specifically, we
% implemented the following two case-studies: {\em (1)} orchestration of
% auto-scaling in micro-services, and {\em (2)} root cause analysis in
% micro-services. For auto-scaling we used ShareLatex~\cite{sharelatex} and for
% root cause analysis we relied on OpenStack~\cite{openstack}. We next describe
% the implementation details of the two case-studies.

In this section, we describe two use cases to demonstrate
\sys's ability to 
handle different workflows. In particular, using \sys's base design,
we 
implemented 1) an orchestration engine for autoscaling and applied it to ShareLatex \cite{sharelatex},
and 2) a root cause analysis (RCA) engine and applied it to OpenStack \cite{openstack}.

%the following two case-studies: (1) orchestration of
%autoscaling; and (2) Root Cause Analysis (RCA). We use two different microservices for the case-studies: ShareLatex~\cite{sharelatex} for autoscaling, and 
%OpenStack~\cite{openstack} for RCA. We next describe the design details of the case-studies based on \sys's design.

\input{case-study-autoscaling}

\input{case-study-rootcause}

%% file: case-study-autoscaling.tex
\subsection{Orchestration of Autoscaling}
\label{subsec:autoscaling}
For the autoscaling case study, 
we used ShareLatex~\cite{sharelatex}---a popular collaborative LaTeX editor. 
ShareLatex is structured as a 
microservices-based application, delegating tasks to multiple well-defined 
components that include a KV-store, load balancer, two databases and 11 node.js based 
components.

\sys's pairwise investigation of representative metrics of components produces the
dependencies across components. 
%These dependencies also cover the time lag when
%one component starts seeing the changes after the first component as well as the
%scaling factor that can be used to add or remove instances of components as
%needed. 
By leveraging this dependency graph, our autoscaling engine helps developers to
make more informed decisions regarding which components and metrics are more
critical to monitor. As a result, developers can generate {\em scaling rules} with the goal of adjusting
the number of active component instances, depending on real-time workload.

More specifically, we use 
\sys's dependency graph and extract 
(1) {\em guiding metrics} (i.e., metrics to use in a scaling rule), (2) {\em scaling actions} (i.e.,
actions associated with reacting to varying loads by increasing/decreasing the number of instances
subject to minimum/maximum thresholds), and (3) {\em scaling conditions} (i.e., 
conditions based on a guiding metric triggering the corresponding scaling action).
Below, we explain how we use \sys to generate a scaling rule:

%A scaling rule is specific to a service and composed by 3 main elements: (1) 
%a `guiding' metric; (2) a scaling action (e.g. scale in\slash out, by 
%how much, minimum and maximum limits for running instances, etc.); and (3) a condition based on the 
%chosen metric, which triggers the scaling action. 

%\include{listings/listings-autoscaling.tex}
%\begin{lstlisting}[label=lst:autoscalingscript,caption=Scaling rule example,
 % frame=single,language=javascript,showstringspaces=false]
% 
%\begin{lstlisting}[label=lst:autoscalingscript,caption=Scaling rule example]
%service web
%metric `http-requests_Project_id_GET_mean'
%action scale_out
%    when(metric > 1400)
%    by(current + 1)
%    min_instances(2)
%    max_instances(10)
%\end{lstlisting}

%Listing~\ref{lst:autoscalingscript} shows an example of an autoscaling rule. It 
%is applied to the \textit{`web'} service, scaling it out by one instance ($+1$), whenever the metric 
%\emph{http-requests\_Project\_id\_GET\_mean} rises over a threshold of 1400\,ms. 
%A `scale in' rule would do the opposite (i.e. $-1$ instance) when the 
%same metric falls below a threshold. Below we explain how we leverage \sys to 
%generate a scaling rule:

\myparagraph{\#1: Metric} We pick a metric $m$ that appears the most
in Granger Causality relations between components.

\myparagraph{\#2: Scaling actions} In our case study, we restrict scaling actions 
to scale in\slash out actions, with increments\slash decrements 
of a single component instance ($+$\slash $-1$). 

\myparagraph{\#3: Conditions} The scale in\slash out thresholds are defined 
from the values of $m$ according to a Service Level Agreement 
(SLA) condition. For ShareLatex, such an SLA condition can be to keep $90\%$ of 
all request latencies below 1000ms. The thresholds for $m$ are iteratively 
refined during the application loading phase.

%% file: case-study-rootcause.tex
\subsection{Root Cause Analysis}
\label{subsec:case-study-rootcause}

%\begin{figure*}[t]
%\centering
%\includegraphics[scale=.425]{figures/rca-flow.pdf}
%\caption{\sys's RCA method.}
%\label{fig:case-study-rootcause-worklow}
%\end{figure*}

% To demonstrate the versatility of \sys, we use the dependency graphs 
% to conduct Root Cause Analysis (RCA) of anomalies in service-oriented 
% applications. Our case study takes Openstack~\cite{openstack} as an example of a 
% service-oriented application. 

% In the next sections, we explain how we conduct 
% root cause analysis of anomalies using dependency graphs, and briefly describe 
% the Openstack application.

% \subsubsection{Overview}

For the root cause analysis (RCA) case study, we used OpenStack~\cite{openstack,openstack-kolla}, a 
popular open-source cloud management software. OpenStack is structured as a microservices-based 
application with a typical deployment of $\sim$10 (or more) individual components, each often 
divided into multiple sub-components~\cite{openstack-hansel}. Due to its scale and complexity, 
OpenStack is susceptible to faults and performance issues, often introduced by updates to its 
codebase. 

% The documentation for such issues is publicly available~\cite{openstack-bugs}, allowing 
% for the identification of ``correct" and ``faulty" version pairs, i.e. code versions 
% which \textit{precede} and \textit{follow} the introduction of the issue, respectively.
% This makes it easier for us to setup scenarios and evaluate the effectiveness of RCA.

%Second, 
%Openstack's service-oriented architecture shares many aspects of a typical microservice
%architecture: its components are loosely coupled, independently updated, and 
%communicate via well-defined APIs~\cite{openstack-hansel}. In fact, several 
%projects are now deploying Openstack as ``containerized microservices'' (e.g. 
%Openstack Kolla~\cite{openstack-kolla}), in some instances going as far as 
%automatically deploying new components from the latest 
%source~\cite{openstack-microservices}.

In microservices-based applications such as Openstack, components can be 
updated quite often~\cite{gremlin}, and such updates can affect other application 
components. If relationships between components are 
complex, such effects 
may not be easily foreseeable, even when inter-component interfaces are unchanged 
(e.g., if the density of inter-component 
relationships is high or if the activation of relationships is selective depending 
on the component's state and inputs).
\sys's dependency graph can be used 
to understand the update's overall effect on the application: changing dependency graphs
can indicate potential problems introduced by an update. By identifying such
changes, \sys can help developers identify the root
cause of the problem. 

Our RCA engine leverages \sys to generate a list 
of possible root causes of an anomaly in the monitored application. More 
specifically, the RCA engine compares the dependency graphs of two different 
versions of an application: (1) a \textit{correct} version; and (2) a  
\textit{faulty} version. Similarly to~\cite{monitorrank, rca-2010}, we assume that the system anomaly 
(but not its \textit{cause}) has been observed and the {correct} and faulty versions have been identified.
The result of this comparison is a list of 
$\{$\textit{component}, \textit{metric list}$\}$ pairs: the \textit{component} item 
points to a component as a possible source for the issue, whereas the \textit{metric list} 
shows the metrics in that component potentially related to the issue, providing a more fine-grained view.
% , electing a group of 
% metrics as related to the root cause. 
With the help of this list, developers can reduce the complexity of their search for 
the root cause.

\begin{table}
%\scriptsize
\myfontsize
\centering
\caption{Description of dependency graph differences considered by the root cause analysis engine.}
\vspace{-2mm}
\label{table:rca-change-types}
\newcommand{\tabitem}{~~~~~\llap{\textbullet}~}
\begin{tabular}{@{}ll@{}}
\toprule
\textbf{Scoping level}              & \textbf{Differences of interest} \\ \midrule
\multirow{2}{*}{Component metrics}    & Present in F version, not in C (\textit{new}) \\
                                    & Present in {C} version, not in F (\textit{discarded}) \\ \midrule
\multirow{1}{*}{Clusters}           & Cluster includes new\slash discarded metrics \\ \midrule
\multirow{3}{*}{Dep. graph edges}   & New\slash discarded edge between similar clusters \\
                                    & Different time-lag between similar clusters \\
                                    & Includes clusters w/ new\slash discarded metrics \\
\bottomrule
\end{tabular}
\end{table}

\begin{figure*}[ht]
\centering
\includegraphics[scale=.45]{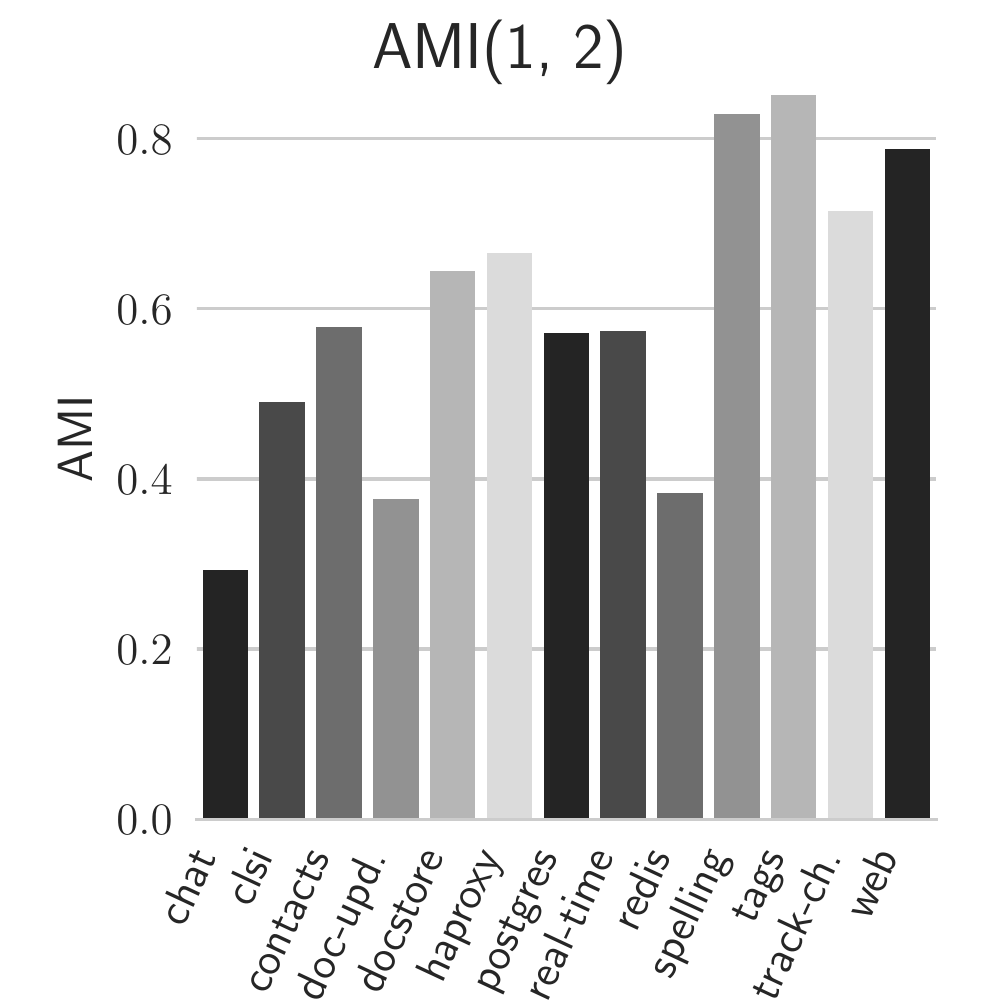}
\includegraphics[scale=.45]{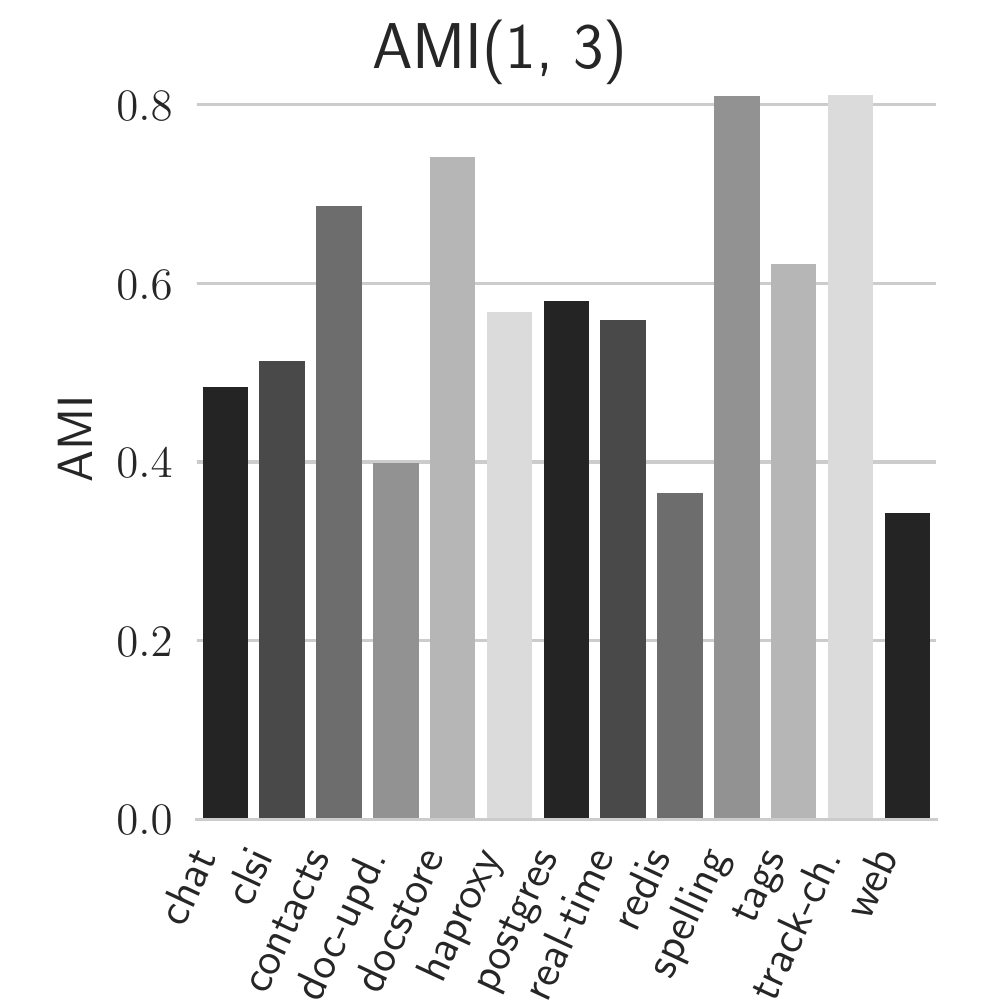}
\includegraphics[scale=.45]{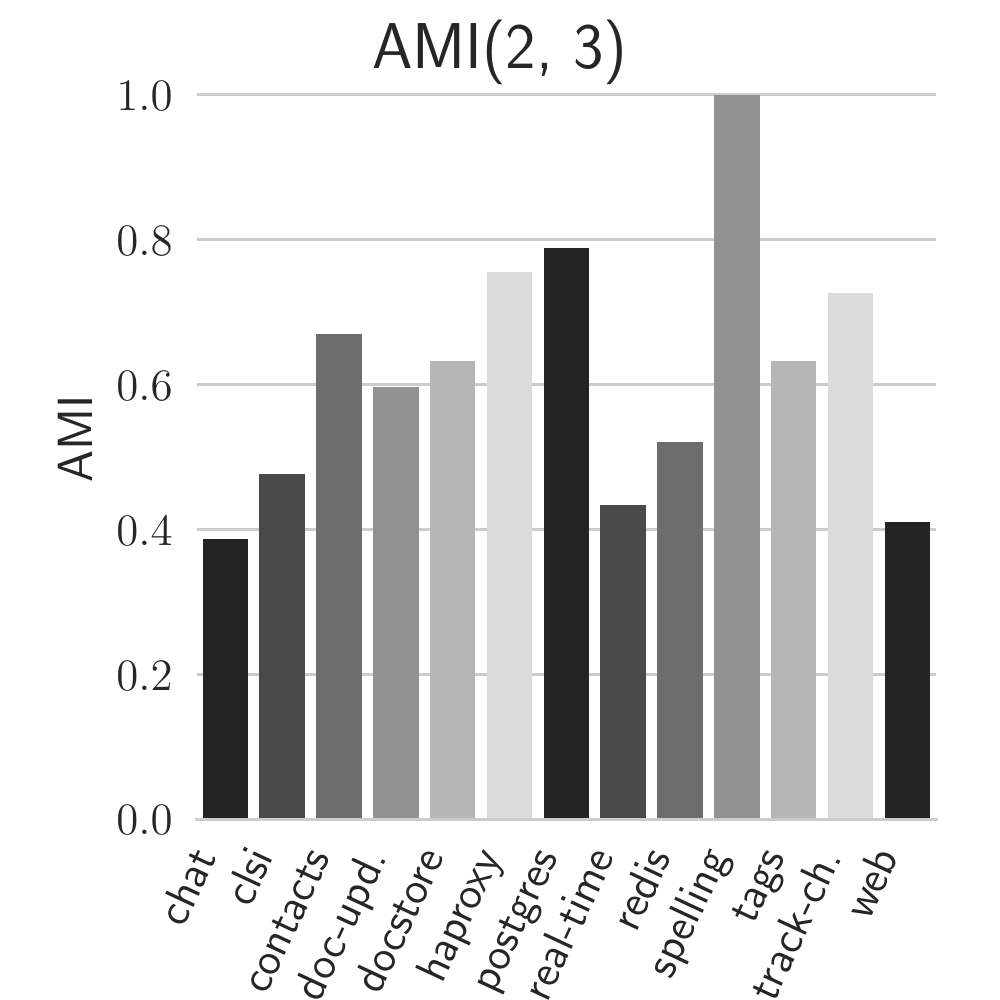}
\caption{Pairwise adjusted mutual information (AMI) scores between 3 measurements.}
\label{fig:mutual-information-score}
\vspace{-3mm}
\end{figure*}

% The basic idea behind the RCA is to compare the dependency graphs (and the clustering metadata) 
% of {the \textit{correct} (C) version} and the \text{faulty} (F) version of the application. 

% , i.e. our RCA method 
% does not detect anomalies on-the-fly. This allows for the identification of 
% non-faulty and faulty versions, over which \sys can run independently.

% The detailed workflow of RCA is as follows: The RCA method relies in the comparison of clustering metadata (i.e. the metrics 
% aggregated within a particular $<$\textit{service, cluster}$>$ pair) and 
% dependency graphs from two different versions of the monitored application. We 
% designate such versions as (1) \textit{non-faulty} (NF) version, in which 
% the anomaly is absent; and (2) \text{faulty} (F), in which the 
% anomaly is present. Similarly to~\cite{monitorrank, rca-2010}, we assume that the system anomaly 
% (but not its \textit{cause}) has been previously detected, i.e. our RCA method 
% does not detect anomalies on-the-fly. This allows for the identification of 
% non-faulty and faulty versions, over which \sys can run independently.

Figure~\ref{fig:case-study-rootcause-worklow} shows the five steps involved in the 
comparison. At each step, we extract and analyze \sys's outputs at three 
different granularity levels: \textit{metrics, clusters}, and {\em dependency graph edges}. 
The levels and corresponding differences of interest are described in Table \ref{table:rca-change-types}. We 
describe the steps in more detail below.

% Figure~\ref{fig:case-study-rootcause-worklow} shows the 
% steps involved in the comparison. At each step, we extract and analyze \sys's outputs at three 
% different scoping levels - \textit{metrics, clusters}, and {\em dependency graph edges} - as 
% summarized in Table~\ref{table:rca-change-types}. 
% The reasoning behind this choice of scoping levels is as follows:

\myparagraph{\#1: Metric analysis} This step analyzes the presence 
or absence of metrics between C and F versions. If a metric $m$ 
is present in both C and F, it intuitively represents the maintenance 
of healthy behavior associated with $m$. As such, these metrics are filtered out of 
this step. Conversely, the appearance of a new metric (or the disappearance of a previously 
existing metric) between versions is likely to be related with 
the anomaly.

\myparagraph{\#2: Component rankings} In this step, we use the results of 
step 1 to rank components according to their \textit{novelty score} (i.e., total number of 
new or discarded metrics), producing an initial group of interesting components for RCA.

\myparagraph{\#3: Cluster analysis: novelty \& similarity} 
Clusters aggregate component metrics which exhibit similar behavior over time. 
The clusters with new or discarded metrics
should be more interesting for RCA compared to the unchanged clusters of that 
component (with some exceptions, explained below). For a given component, we compute the novelty 
scores of its clusters as the sum of the number of new and discarded metrics, and produce a list of 
$\{$\textit{component}, \textit{metric list}$\}$ pairs, where the metric list considers metrics from the
clusters with higher novelty scores.
% the metrics aggregated in clusters containing new 
% and/or discarded metrics are interesting for RCA, leading us to consider 
% \novelty scores \textit{of the cluster}, the more interesting it should be.
% This scoping level refines the initial group of services selected in the previous step, producing 
% a list of $\{$\textit{service}, \textit{list(metric)}$\}$ pairs.

% why do we need the clusters? 
% 1) they are the output of sieve, so we have to stick with that...
% 2) they they identify metrics which present a similar behavior over time. 

In addition, we track the similarity of a component's clusters between {C} and F 
versions (or vice-versa). This is done to identify two events: (1) appearance (or disappearance) of 
edges between versions; and (2) attribute changes 
in relationships maintained between C and F versions (e.g., a change in Granger causality time lag). 
An edge between clusters $x$ and $y$ (belonging to components $A$ and $B$, respectively) 
is said to be `maintained between versions' if their respective metric 
compositions do not change significantly between C and 
F versions, i.e. if $S(\mathcal{M}^{A}_{x,\text{C}}) \approx S(\mathcal{M}^{A}_{x',\text{F}})$ \textbf{and} $S(\mathcal{M}^{B}_{y,\text{C}}) \approx S(\mathcal{M}^{B}_{y',\text{F}})$. {$\mathcal{M}^{A}_{x,\text{C}}$ and $\mathcal{M}^{A}_{x',\text{F}}$} are the metric 
compositions of clusters $x$ and $x'$ of component $A$, in the 
{C} and F versions, respectively. $S$ is some measure of cluster similarity 
(defined below). Both events -- (1) and (2) -- can be an indication of an anomaly, 
because one would expect edges between clusters with high similarity 
to be maintained between versions.

% $\mathcal{M}^{A}_{c,\text{C}}$
% $\mathcal{M}^{A}_{c,\text{F}}$

% $\mathcal{M}^{B}_{d,\text{C}}$
% $\mathcal{M}^{B}_{d,\text{F}}$

% The attributes of cluster relationships can also change; for example, if 
% the Granger Causality time-lag increases between versions, 
% the relationship and associated clusters can be linked to a latency anomaly 
% in the system.
We compute the \textit{cluster similarity score}, $S$, according to a modified form of the 
Jaccard similarity coefficient
\begin{align}
	S = \frac{|\mathcal{M}^{A}_{i,\text{C}} \cap \mathcal{M}^{A}_{j,\text{F}}|}{|\mathcal{M}^{A}_{i,\text{C}}|}
\end{align}
% where {$\mathcal{M}^{A}_{i,\text{C}}$ and $\mathcal{M}^{A}_{j,\text{F}}$} are the metric 
% compositions of clusters $i$ and $j$, belonging to some component $a$, in the 
% {C} and F application versions. 
To eliminate the penalty imposed by 
new metrics added to the faulty cluster, we only 
consider the contents of the {correct} cluster 
in the denominator (instead of the union of {$\mathcal{M}^{A}_{i,\text{C}}$ 
and $\mathcal{M}^{A}_{j,\text{F}}$}).

\myparagraph{\#4: Edge filtering} To further reduce the list of 
$\{$\textit{component}, \textit{metric list}$\}$ pairs, we examine the 
relationships between components and clusters identified in steps 2 and 3. We 
identify three events:

\begin{enumerate}
\item Edges involving (at least) one cluster with a high novelty score
\item Appearance or disappearance of edges between clusters with high similarity 
\item Changes in time lag in edges between clusters with high similarity
\end{enumerate}

Event 1 isolates metrics related to edges which include at least one 
`novel' cluster. Events 2 and 3 isolate clusters 
which are maintained between C and F versions, but become 
interesting for RCA due to a change in their relationship. Novelty 
and similarity scores are computed as in step 3. We define 
thresholds for `high' novelty and similarity scores.

\myparagraph{\#5: Final rankings} We present a final list 
of $\{$\textit{component}, \textit{metric list}$\}$ pairs. The list 
is ordered by component, following the rank given in step 2. The \textit{metric list} 
items include the metrics identified at steps 3 and 4.

%% file: implementation.tex
%\begin{figure*}[ht]
%\centering
%\includegraphics[scale=.45]{figures/mutual-information-score-1-2.pdf}
%\includegraphics[scale=.45]{figures/mutual-information-score-1-3.pdf}
%\includegraphics[scale=.45]{figures/mutual-information-score-2-3.pdf}
%\caption{Pairwise adjusted mutual information (AMI) scores between 3 measurements.}
%\label{fig:mutual-information-score}
%\vspace*{-0.1cm}
%\end{figure*}

\section{Implementation}
\label{sec:implementation}

We next describe the implementation details of \sys. 
Our system implementation, including used software
versions, is published at \url{https://sieve-microservices.github.io}.
For load generation, \sys requires an application-specific load 
generator. We experimented with two microservices-based applications: ShareLatex~\cite{sharelatex} and OpenStack~\cite{openstack, openstack-kolla}. For ShareLatex, we developed our own load generator using Locust \cite{locust}, a Python-based distributed load generation tool to simulate
virtual users in the application ($1,041$ LoC). For OpenStack, we used Rally \cite{openstack-rally}, 
the official benchmark suite from OpenStack.

For metric collection, \sys uses Telegraf \cite{telegraf} to collect
application/system metrics and stores them in InfluxDB \cite{influxdb}. Telegraf seamlessly
integrates with InfluxDB, supports metrics of commonly-used components (e.g., Docker, RabbitMQ, memcached)
and can run custom scripts for collection of additional metrics exposed by application 
APIs (e.g., \cite{openstack-telemetry}). With this setup, \sys can store any time-series metrics
exposed by microservice components.

For the call graph extraction, \sys leverages sysdig call tracer~\cite{sysdig} to obtain which
microservice components communicate with each other. We wrote custom 
scripts to record network system calls with source and destination IP addresses 
on every machine hosting the components ($457$ LoC). These IP addresses are then mapped to the components
using the cluster manager's service discovery mechanism.

We implemented \sys's data analytics techniques in Python ($2243$ LoC) including metric filtering, clustering based on \kshape, and Granger Causality.  
The analysis can also be distributed across multiple machines for scalability.

Lastly, we also implemented two case studies based on the \sys 
infrastructure: autoscaling in ShareLatex (720 LoC) and RCA in 
OpenStack ($507$ LoC).  For our autoscaling engine, %we employ a time-series database called InfluxDB \cite{influxdb}
%and feed it with observed metrics data using Telegraf \cite{telegraf}. 
we employed Kapacitor \cite{kapacitor} to stream metrics from InfluxDB in real-time 
and to install our scaling rules using its user-defined functions. For the RCA 
engine, we implemented two modules in Python: one module extracts metric clustering data ($125$ LoC)
and the other module ($382$ LoC) compares clustering data and dependency graphs. % produced by \sys.

%% file: evaluation.tex
\section{Evaluation}
\label{sec:evaluation}

%In this section, we first describe the general implementation of \sys
%(\S \ref{sec:implementation}). 
%We then present the evaluation of \sys aiming to answer the following 
%questions:
Our evaluation answers the following questions: 
\begin{enumerate}
\item How effective is the general  \sys framework? ($\S$\ref{subsec:sieve-evaluation})
\item How effective is \sys for autoscaling? ($\S$\ref{subsec:auto-scaling-evaluation})
\item How effective is \sys for root cause analysis? ($\S$\ref{subsec:root-cause-evaluation})
\end{enumerate}

%Before answering these evaluation questions, we briefly described the implementation details.

\input{eval-general}

\input{eval-auto-scaling}

\input{eval-root-cause}

%% file: eval-general.tex
\subsection{Sieve Evaluation}
\label{subsec:sieve-evaluation}

Before we evaluate \sys with the case studies, we evaluate \sys's general properties: {\em (a)} the robustness of clustering; {\em (b)} the effectiveness of metric reduction; and {\em (c)} the monitoring overhead incurred by \sys's infrastructure.

\myparagraph{Experimental setup}  We ran our measurements on a 10 node cluster,
every node with a 4-core Xeon E5405 processor, 8 GB DDR2-RAM and a 500GB HDD.
For the general experiments, we loaded ShareLatex using \sys five times with
random workloads. The random workloads also help to validate whether the model stays consistent, if no assumption about the workload is made.

\subsubsection{Robustness}
\label{subsubsec:robustness}
We focus on two aspects to evaluate \sys's robustness. First, we investigate the {\em consistency} of clustering across different runs. Second, we try to {\em validate} whether the metrics in a cluster indeed belong together.

%First, we try to understand how
%{\em consistent} the clustering across different runs is.
%%In other words, we would like to see whether the clustering
%%is going to produce different clusters at  different runs.
%

%. In other words,
%we would like to validate that the metrics in a cluster really belong together.

\myparagraph{Consistency} To evaluate consistency, we compare cluster assignments
produced in different measurements. 
A common metric to compare cluster assignments is 
Adjusted Mutual Information (AMI) score~\cite{informationmeasures}. AMI is normalized against a random assignment and ranges from zero to one: 
If AMI is equal to one, both clusters match perfectly. Random assignments will be close to zero.

%AMI is an extension of 
%mutual information ($I$), which is defined as:
%\begin{align}
%I(X, Y) &= H(X) - H(X|Y) \\
%          &= H(Y) - H(Y|X)
%\end{align}
%where $H$ represents the entropy in bits.
%Namely, MI quantifies the entropy obtained from one variable $X$ if the other
%variable $Y$ is already known. AMI is defined as follows:
%\begin{align}
%  AMI(X, Y) &= \frac{I(X,Y)  - E(I(X,Y))} {max(H(X), H(Y)) - E(I(X, Y))}
%\end{align}
%where $E(I(X,Y))$ represents the expected value of the mutual information between $X$ and $Y$. $AMI$ is normalized against a random assignment and ranges from zero to one: 
%If AMI is equal to one, both clusters match perfectly. Random assignments will be close to zero.

Figure~\ref{fig:mutual-information-score} shows the AMI of cluster assignments for individual components for three
independent measurements. To reduce the selection bias we apply randomized workload in a controlled environment. As a result, they should constitute a worst-case performance for the clustering. Our measurements show that the average AMI is 0.597, which is better than random assignments.
%The score tends to get higher for those services, which have clusters with higher
%Silhouette scores (i.e., higher clustering quality). 
Based on these measurements, we conclude the clusterings are consistent.

\begin{figure}
\center
\includegraphics[scale=.35]{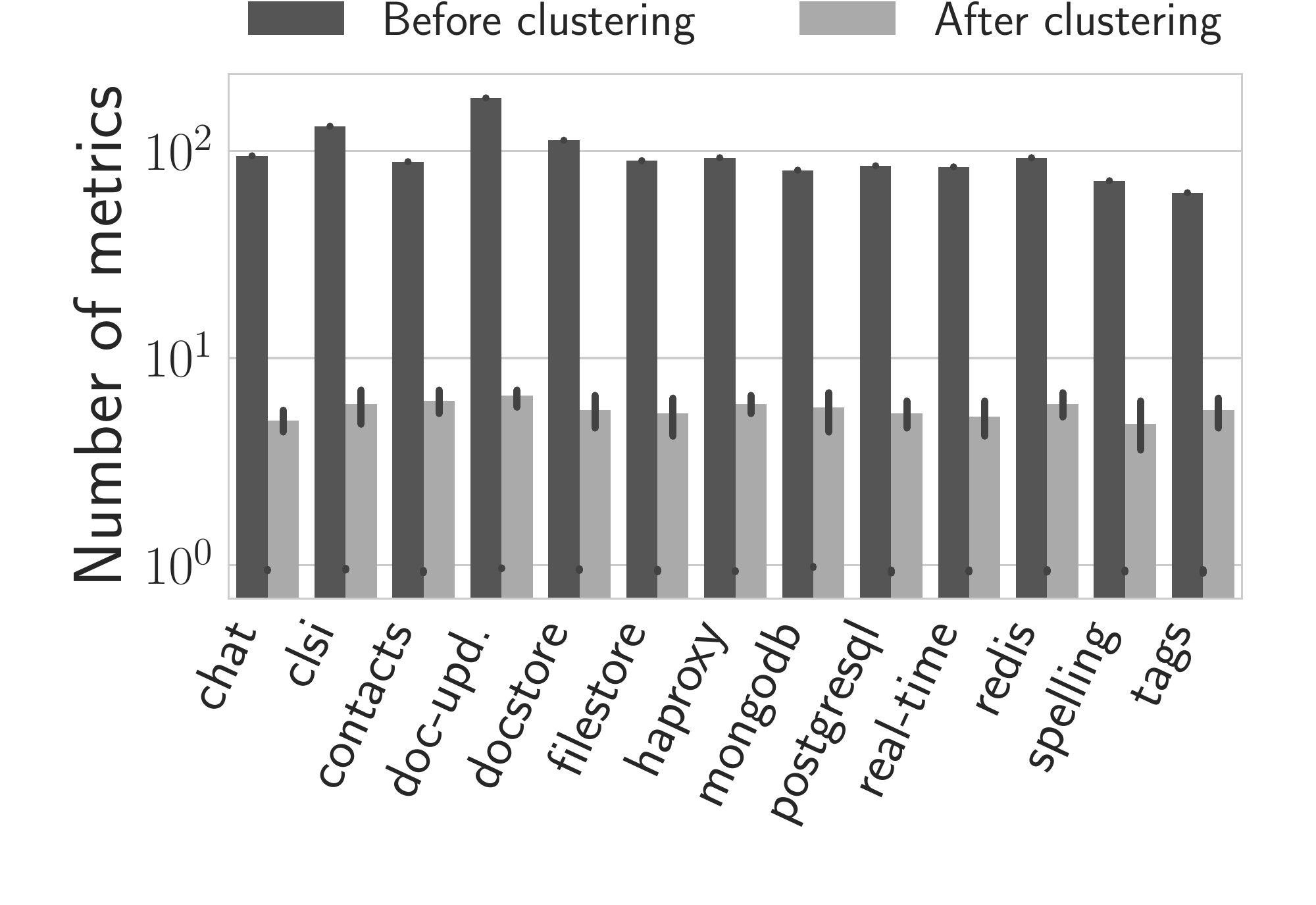}
\vspace*{-0.5cm}
\caption{Average no. of metrics after \sys's reduction.}
\label{fig:metric-reduction}
\end{figure}

\myparagraph{Validity} To evaluate the validity of the clusters, we choose three criteria: 
\emph{(1)} Is there a common visible pattern between metrics in one cluster?
\emph{(2)} Do metrics in a cluster belong together assuming application knowledge?
\emph{(3)} Are the shape-based distances between metrics and their cluster centroid below a threshold (i.e., 0.3)?

We choose three clusters with different Silhouette scores (high, medium, low).
According to the above criteria, we conclude the clustering algorithm can determine 
similar metrics. For example, application metrics such as HTTP request times and corresponding
database queries are clustered together. Similar to consistency, higher Silhouette scores indicate
that the clusters are more meaningful and potentially more useful for the developers. 
We omit the details for brevity.

\subsubsection{Effectiveness}
\label{subsubsec:effectiveness}

The purpose of clustering is to reduce the number of metrics exposed by the system
without losing much information about the system behavior.  To evaluate how effective our clustering is in reducing the number of metrics, we
compare the results of the clustering with the actual number of metrics in the application.
We identified 889 unique metrics within ShareLatex, meaning that 
an operator would have to understand and filter these metrics. \sys's clustering reduces this
number to 65 (averaged across five runs). Figure \ref{fig:metric-reduction} shows the
reduction in the number of metrics for the individual components in ShareLatex. Note that this measurement is with high Silhouette scores for the clusters, which implies that the metrics reduction does not affect the quality of the clusters.

%\pramod{It doesn't say much the quality of clustering? Should we mention the silhouttee index here for it?}

%The goal of clustering metrics was to reduce the number of metrics exposed by the system
%to a smaller set which is suitable to describe the system. From each cluster the
%one metric with closest distance to its centroid is picked. As distance measure
%we use the Shape Based distance, which was the measure used to compute the
%clustering (see Section~\ref{sec:design:metric_reduction}).
%That way we can assure the best real-world representative of our cluster is
%picked as the centroid is mean of all metrics.

% After the clustering, we could reduce
%To evaluate how many metrics can be reduced, we run random workload five times
%for one hour each. The resulting metric time series were then clustered using
%\kshape~\ref{sec:design:metric_reduction}. In figure~\ref{fig:metric-reduction} we see the
%number metrics before and after clustering for individual services. We
%identified 889 unique metrics across the application. After clustering we are
%able to reduce this number to 65 on average.

\subsubsection{Monitoring Overhead}
\label{subsubsec:monitoring}
We evaluate \sys's overhead based on two aspects. First, we compare different techniques for obtaining
the call graph and show how our approach fairs. Second, we investigate the overhead
incurred by the application by comparing the monitoring overhead with and without using
\sys.

\begin{figure}
\center
\includegraphics[scale=.28]{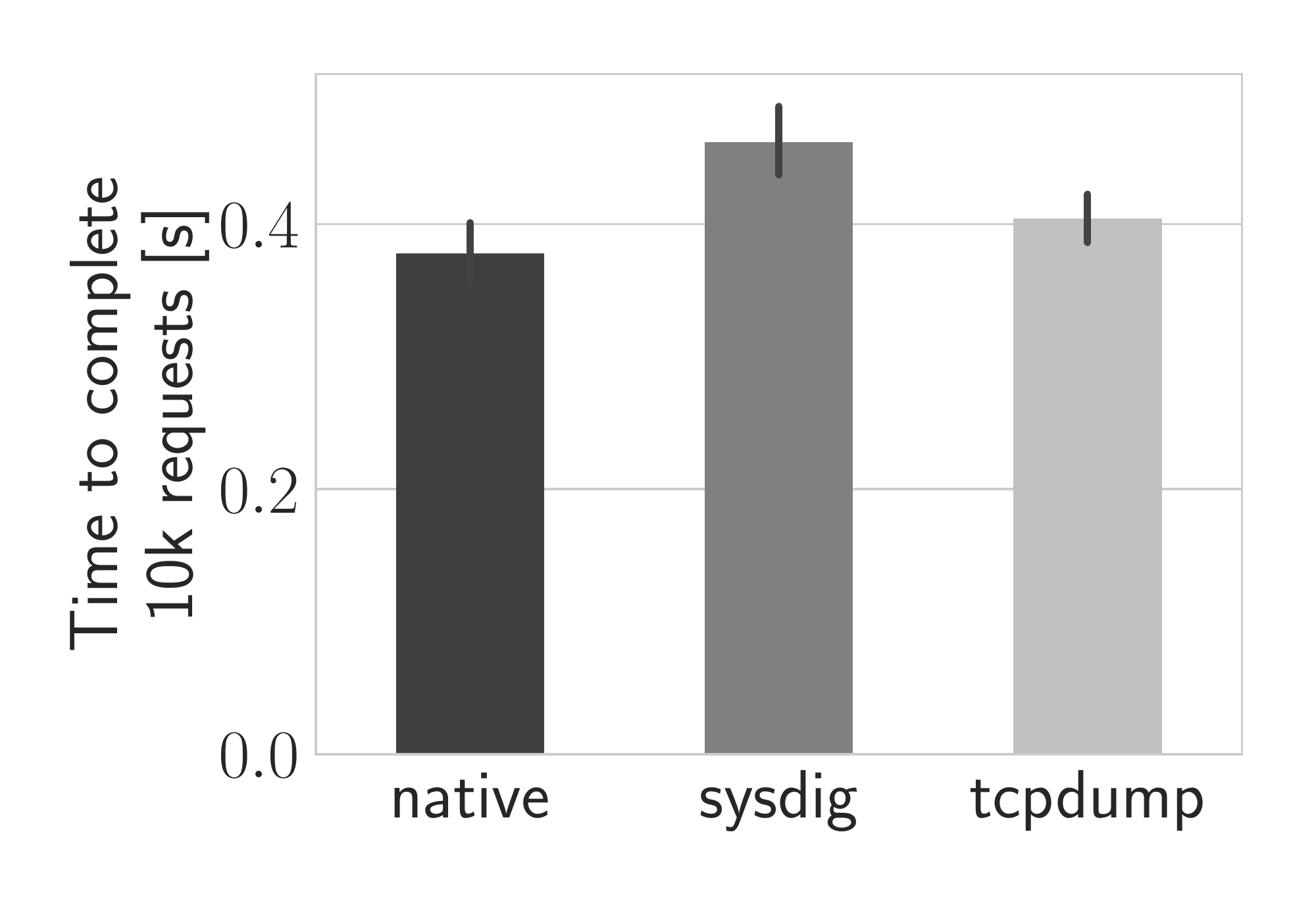}
\caption{Completion time for HTTP requests when using tcpdump, sysdig or native (i.e., no monitoring).}
\label{fig:time-overhead-sysdig}
\end{figure}

\myparagraph{Overheads} To measure the monitoring overhead during the loading stage,
we run an experiment with 10K HTTP requests for a small static file using Apache Benchmark
\cite{apachebenchmark} on an Nginx web server \cite{nginx}.
Because the computational overhead for serving such a file is low, this experiment
shows the worst-case performance for sysdig and tcpdump.
Figure \ref{fig:time-overhead-sysdig} shows the time it takes to complete the experiment.
While tcpdump incurs a lower overhead than sysdig (i.e., 7\% vs. 22\%), it provides
less context regarding the component generating the request and requires more knowledge
about the network topology to obtain the call graph. sysdig provides all this information
without much additional overhead.

% To obtain the callgraph from section~\ref{ssec:callgraph} it is required to
% trace process communication using sysdig. In a microbenchmark we wanted to
% evaluate what overhead it adds compared to tcpdump or execution without tracing.
% In figure~\ref{fig:time-overhead-sysdig} we measured  the time it took to complete
% 10K HTTP Requests of a small static file by Apache Benchmark against the
% webserver Nginx. As computational overhead in both applications for this type of
% request is rather low the benchmark shows the worst case performance of both
% approaches. We observed an overhead of 22 percent for sysdig and a seven percent
% overhead when using tcpdump. Tcpdump introduces a lower overhead,
% but also provides less context which process in which application container
% issues the request. It also requires more knowledge about the network topology to
% compute the callgraph in case when Overlay networks and Network Address
% Translation (NAT) is used.

\myparagraph{Gains} To show the gains during the runtime of the application after using \sys, we compare the
computation, storage and network usage for the metrics collected during the five 
measurements. We store all collected metrics in InfluxDB and measure the respective resource usage. 
We then repeat the same process using the metrics found by \sys; thus, simulating a run
with the reduced metrics.
Table \ref{fig:influxdb-overhead} shows the relative usage of the respective
resources with \sys. \sys reduces the monitoring overhead for computation, storage and network
by 80\%, 90\% and 50\%, respectively.

% if less metrics are required the overhead of monitoring services can be
% reduced.  In our monitoring setup we used InfluxDB to collect time series.  To
% measure the storage and computation overhead saved, we reinserted our previous
% measurements with full and reduced set of metrics. We compared cpu time,
% network bandwidth and database size for five measurements. In
% Figure~\ref{fig:influxdb-overhead} we see a reduction of over 80\% for CPU
% time, a 90\% reduction for the resulting database size as well as 40\% to 50\%
% percent reduction in network traffic.

\begin{table}
\centering
%\scriptsize
\caption{InfluxDB overhead before \sys's reduction of metrics.}
\vspace{-2mm}
\myfontsize
\newcommand{\tabitem}{~~~~~\llap{\textbullet}~}
\setlength{\tabcolsep}{2pt}
\begin{tabular}[b]{lllcl}

\toprule
\textbf{Metric} &   \textbf{Before} &  \textbf{After} & \textbf{Reduction} \\
\midrule
CPU time [s]        &      0.45G &    0.085G  & 81.2 \% \\
DB size [KB]        &      588.8 &       36.0  & 93.8 \% \\
Network in [MB]      &       11.1 &        2.3  & 79.3 \% \\
Network out [KB]      &       15.1 &        7.4  & 50.7 \% \\
\bottomrule
\end{tabular}

\label{fig:influxdb-overhead}
\vspace{-3mm}
\end{table}

%% file: eval-auto-scaling.tex
\subsection{Case-study \#1: Autoscaling}
\label{subsec:auto-scaling-evaluation}

We next evaluate the effectiveness of \sys for the orchestration of autoscaling
in microservices.

\myparagraph{Experimental setup}
For the autoscaling case study,  we used ShareLatex~\cite{sharelatex} (as
described in $\S$\ref{subsec:autoscaling}). We used $12$ t2.large VM-Instances
on Amazon EC2 with 2 vCPUs, 8GB RAM and 20 GB Amazon EBS storage.
This number of instances were sufficient to stress-test all components of
the application.
The VM instances were allocated statically during experiments as Docker containers. We
created a Docker image for each ShareLatex component and used Rancher
\cite{rancher} as the cluster manager to deploy our containers across different
hosts.

\myparagraph{Dataset} We used a HTTP
trace sample from soccer world cup 1998~\cite{worldcup98} for an hour long trace.
Note that  the access pattern and requested resources in the world cup trace
differs from the ShareLatex application.  However, we used the trace to map
traffic patterns for our application to generate a realistic spike workload. In
particular, sessions in the HTTP trace were identified by using the client IP.
Afterwards, we enqueued the sessions based on their timestamp, where a virtual
user was spawned for the duration of each session and then stopped.

\newcommand{\httpmetric}{\emph{http-requests\-\_Project\-\_id\-\_GET\-\_mean}}

\begin{figure}[t]
\centering
\includegraphics[scale=.3,trim=0cm 0cm 0cm 7cm,clip]{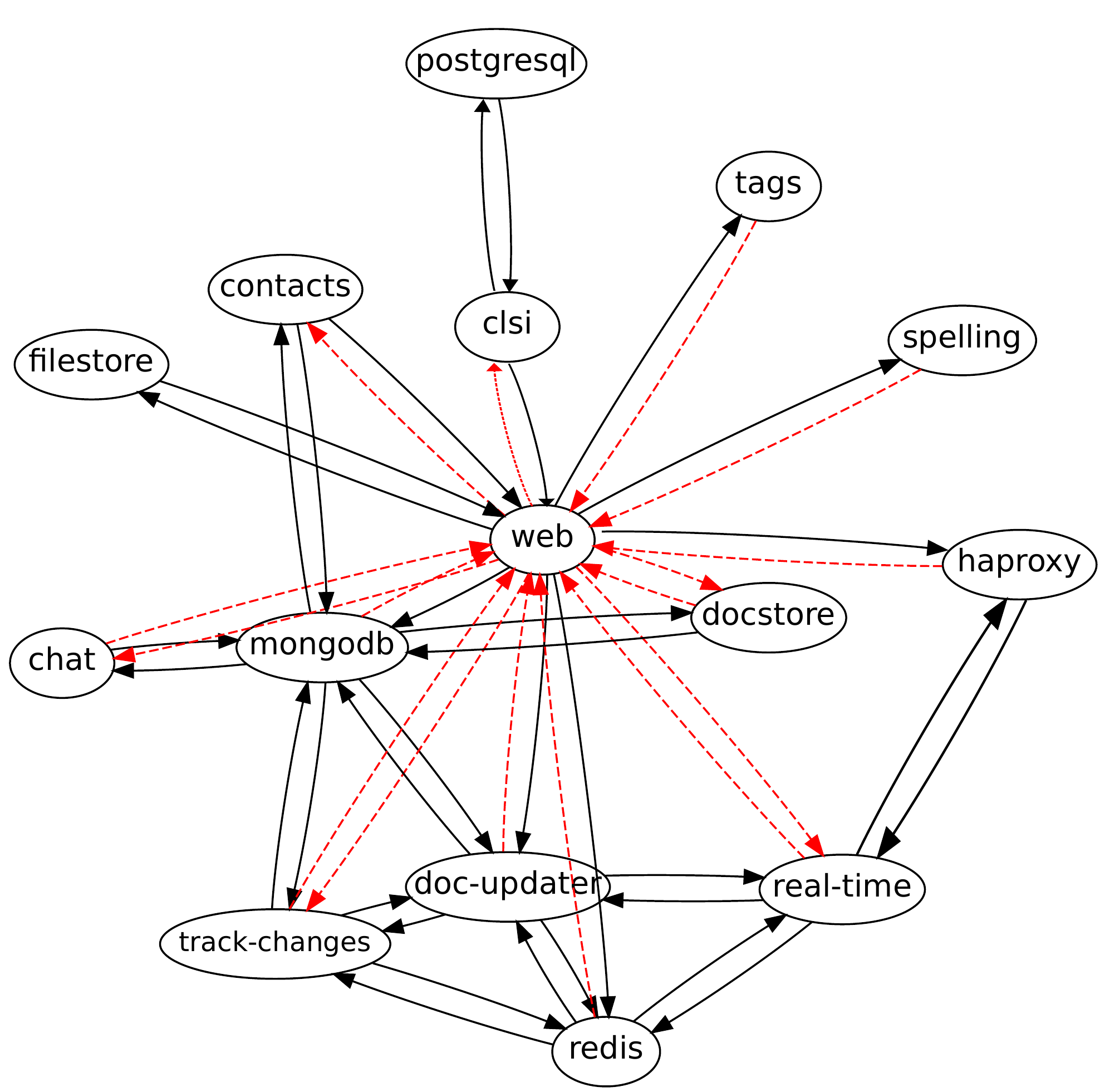}
\caption{Relations between components based on Granger Causality in
	ShareLatex. The dashed lines denote relationships with metric
	http-requests\_Project\_id\_GET\_mean.}
%	, whereas the black lines denote
%	other services relations.}
\label{fig:autoscaling-graph}
\end{figure}

\myparagraph{Results} We chose an SLA condition, 
such that $90th$ percentile of all request latencies should be below 1000ms. 
Traditional tools, such as Amazon AWS Auto Scaling~\cite{awsautoscaling}, often use the
CPU usage as the default metric to trigger autoscaling. \sys
identified an application metric named \httpmetric{} (Figure \ref{fig:autoscaling-graph}) 
as a better metric for autoscaling than CPU usage.

To calculate the threshold values to trigger autoscaling, we used a 5-minute sample
from the peak load of our HTTP trace and iteratively refined the values to stay
within the SLA condition.
As a result, we found that the trigger thresholds for scaling up and down 
while using the CPU usage metric should be 21\% and 1\%, respectively. Similarly,
for \httpmetric{}, the thresholds for scaling up and down should be 1400ms and 1120ms, respectively.

% We found that in case of CPU usage, the scaling engine
% scales out if its mean is over $21$\% and scales down if the mean is below
% $1\%$. Whereas, the application metric  (\httpmetric) identified by \sys scales
% out if the mean is over 1400ms and scales down if the mean is below 1120ms.
% These thresholds were calculated by applying a 5 minute sample of the peak load of our
% HTTP trace iteratively.

After installing the scaling actions, we ran our one-hour trace. 
Table \ref{fig:autoscaling-comparison} shows the comparison when using the CPU usage
and \httpmetric{} for the scaling triggers.
When \sys's selection of metric was used for autoscaling triggers, the average CPU usage of each component
was increased. There were also fewer SLA violations and scaling actions.

\begin{table}
\caption{Comparison between a traditional metric (CPU usage) and \sys's selection when used as autoscaling triggers.}
\vspace{-2mm}
\myfontsize
\begin{tabular}{p{3cm}rrr}
\toprule
\textbf{Metric} & \textbf{CPU usage} & \textbf{Sieve} & \textbf{Difference [\%]} \\
\midrule
Mean CPU usage per component           & 5.98 &          9.26 &         +54.82\\
SLA violations (out of 1400 samples) &  188 &            70 &        -62.77\\
Number of scaling actions           &   32 &            21 &         -34.38\\
\bottomrule
\end{tabular}

\label{fig:autoscaling-comparison}
\vspace{-3mm}
\end{table}

%% file: eval-root-cause.tex
\subsection{Case-study \#2: Root Cause Analysis}
\label{subsec:root-cause-evaluation}

\begin{figure*}[!tp]
\centering
\includegraphics[scale=.4]{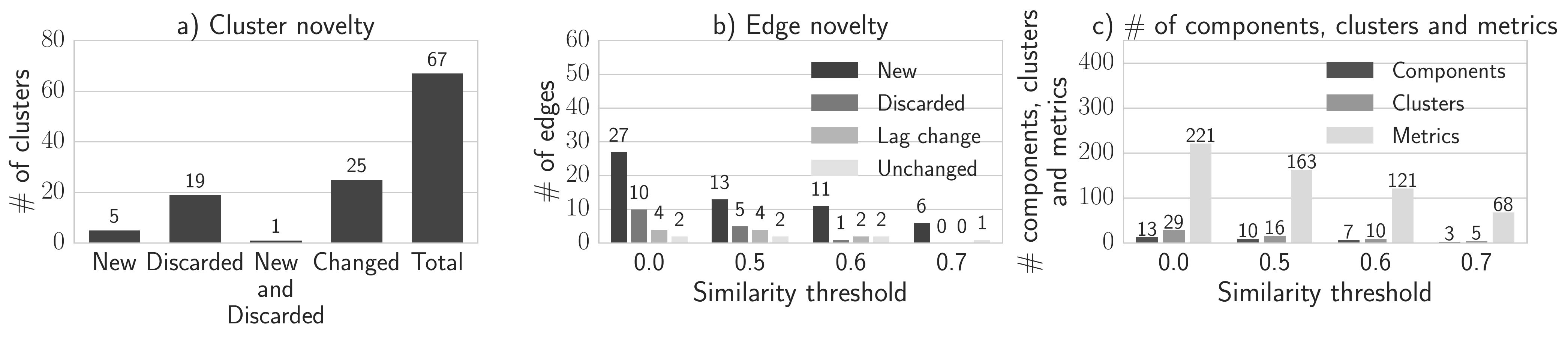}
\vspace*{-.3cm}
\caption{{\bf (a)} Cluster novelty score. {\bf (b)} Edge novelty score.  {\bf (c)} No. of components \& 
clusters after edge filtering w/ varying thresholds.}
\vspace*{-.3cm}
\label{fig:rca-evaluation-clusters}
\end{figure*}

\begin{figure}[!tp]
\centering
\includegraphics[scale=.35]{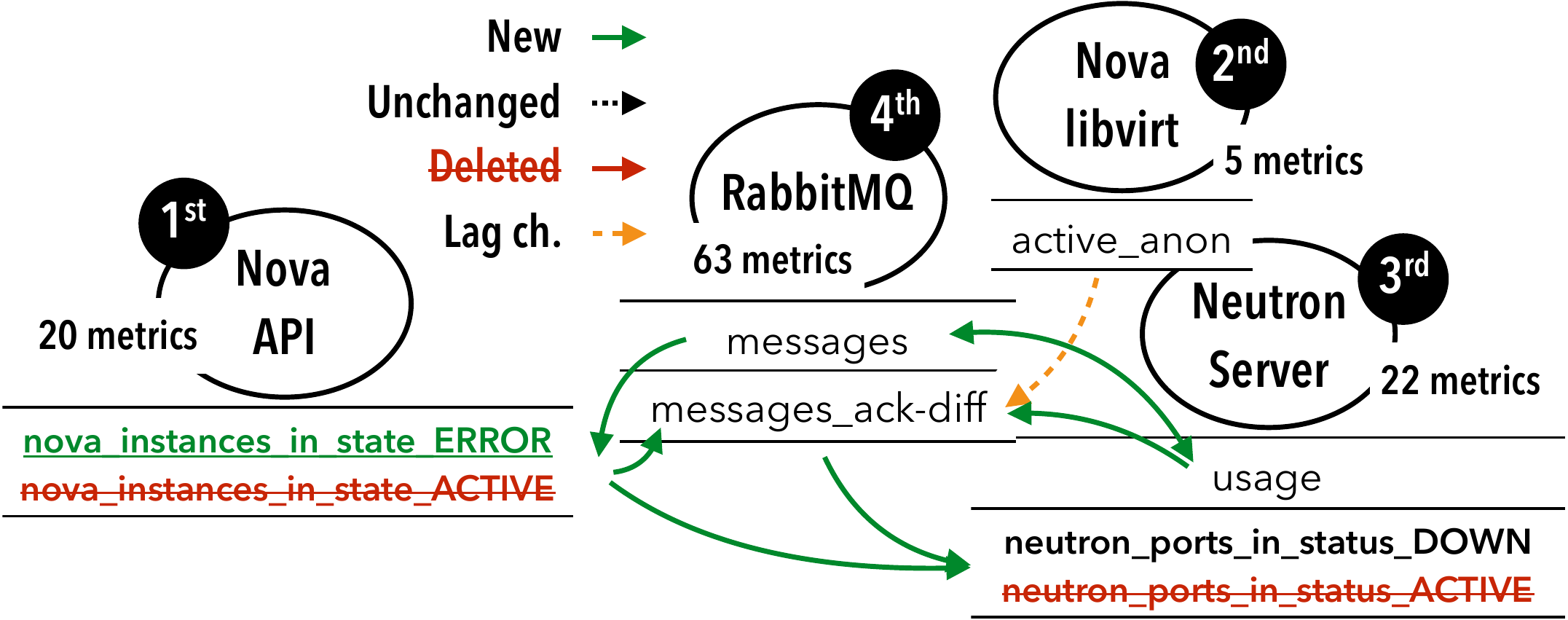}

\caption{Final edge differences for RCA evaluation between top 5 components of 
Table~\ref{table:rca-eval-individual-metrics} with similarity threshold of 
0.50.}
%Note that one of the top 5 components - Nova scheduler - has been 
%discarded by similarity filtering.}

\label{fig:rca-evaluation-edge-differences}
\end{figure}

To evaluate the applicability of \sys to root cause analysis, we reproduce two representative
OpenStack anomalies, Launchpad bugs \#1533942~\cite{launchpad-openvswitch-bug} and 
\#1590179~\cite{launchpad-bug-1590179}. 
{We selected these issues because they have well-documented root causes, providing an 
appropriate ground truth, and allowing for 
the identification of `correct' and `faulty' code versions. Moreover, 
these showcase \sys's effectiveness while analyzing two different types of bugs: 
(1) a crash in the case of \#1533942; and (2) performance regression (e.g. 
increase in latency) in the case of \#1590179.
} We compare the documented root causes to the lists of root causes produced
by our RCA engine.

% explain the bugs
% explain the rankings
% related work

\subsubsection{Bug 1533942 : Failure to launch a VM}
\label{subsubsec:root-cause-evaluation-1533942}

% In this section, we describe x representative OpenStack issues, including
% their root causes.
% As a representative example, we picked 
Bug \#1533942 manifests itself as follows: when launching a new VM instance 
using the command line interface, one gets the error 
message \textit{`No valid host was found. There are not enough hosts available.'} 
despite the availability of compute nodes. Without any other directly 
observable output, the instance falls into `ERROR' state and fails. A similar failure is used as a representative 
case in~\cite{openstack-hansel, openstack-gretel}. 

\myparagraph{Root cause} The failure is caused by the crash of an agent in the 
Neutron component, namely the Open vSwitch agent. The Open vSwitch agent is responsible for 
setting up and managing virtual networking for VM instances. The ultimate cause is traced 
to a configuration error in OpenStack Kolla's deployment scripts \cite{launchpad-openvswitch-bug}.

\myparagraph{Experimental setup} We deployed OpenStack components as 
containerized microservices using Kolla~\cite{openstack-kolla}. We 
configured Kolla to deploy 7 main OpenStack components along with 12 auxiliary 
components. Some components can be sub-divided in several microservices and replicated among 
deployment hosts, for a total of 47 microservices. This information 
is summarized in Table~\ref{table:rca-eval-components}. We use OpenStack's telemetry component (Ceilometer) 
to expose relevant OpenStack-related metrics and extract them via Telegraf.

The Openstack versions used for the correct (C) and faulty (F) versions are listed 
in Table~\ref{table:rca-eval-openstack-versions}. The configurations for the Kolla deployments of 
each version are publicly available~\footnote{\href{https://github.com/sieve-microservices/kolla}{https://github.com/sieve-microservices/kolla}}, 
as well as the monitoring infrastructure and evaluation scripts~\footnote{\href{https://github.com/sieve-microservices/rca-evaluation}{https://github.com/sieve-microservices/rca-evaluation}}.

The infrastructure consists of two m4.xlarge Amazon EC2 VM instances to run OpenStack components (16 vCPUs, 64 GB RAM and 
20 GB Amazon EBS storage) and three t2.medium VM instances (2 vCPUs, 4GB RAM and 
20 GB EBS storage) for the supporting components (measurement, database and deployment).

\myparagraph{Results} We expect the RCA engine's outcome to include 
Neutron component, along with metrics relating VM launches and networking.
% whose relationships relate VM launch and VM networking anomaly. 
The $\{$\textit{component}, \textit{metrics list}$\}$ pairs with Neutron should be ranked higher than others.

To generate load on OpenStack, we run the `boot\_and\_delete' (B\&D) task 100 times with 
the Rally benchmark suite \cite{openstack-rally}, 
which launches 5 VMs concurrently and deletes them after 15-25 seconds (details 
about Rally tasks in Table~\ref{table:rca-eval-rally-tasks}). We apply this 
process to the {correct (C)} and faulty (F) versions of OpenStack. For the faulty version, the task fails as 
described above. We then apply the remaining stages of \sys and feed the output 
to the RCA engine. For both versions, the dependency graphs are composed 
by 16 components, with 647 edges in the NF version, and 343 edges 
in the F version. Below, we summarize the findings of RCA steps.

\myparagraph{Steps \#1 \& \#2: Metric analysis and component rankings} The total 
number of unchanged metrics exceeds that of `novel' metrics (i.e., new and/or 
discarded) by an order of magnitude. Furthermore, the initial 
component novelty ranking puts the Nova and Neutron components (known 
to be directly related with the anomaly) within the top 4 
positions out of 16 (Table \ref{table:rca-eval-individual-metrics}). 
% \iea{@antonio: can you modify the table to show the total (i.e., as novelty score maybe?) and update the table's caption to say its sorted accordingly? you can use numbers of changed and unchanged
% metrics inside parentheses. for example, Nova API 29(7/22) 30}
This confirms the intuition behind our approach: novel metrics are more 
likely to be related to a failure.
% \iea{the following seems redundant, because we said the same thing just above}
% because the anomalies are 
% directly related to observable failures in VM management (Nova) and hidden 
% failures in VM networking (Neutron).

%\begin{figure*}[!ht]
%
%  \subfloat[Cluster and Edge novelty scores (a and b). Number of components and 
%clusters after edge filtering with diff. similarity thresholds (c).\label{fig:rca-evaluation-clusters}]{%
%    \includegraphics[scale=.4]{figures/rca-evaluation-clusters.pdf}
%  }
%%   \quad
%%  \subfloat[Final edge differences for RCA evaluation, top 5 components, similarity threshold of 
%%0.50\label{fig:rca-evaluation-edge-differences}]{%
%%    \includegraphics[scale=.35]{figures/rca-edge-diff-alt.pdf}
%%  }
%  \caption{Results from Root Cause Analysis (RCA) case-study}
%  \label{fig:overhead-evaluation}
%\end{figure*}

\myparagraph{Step \#3: Cluster novelty \& similarity} 
Computing the cluster novelty scores shows that the novel metrics from step 1
are distributed 
over only 27 of the 67 clusters (Figure \ref{fig:rca-evaluation-clusters}(a)), even
conservatively considering a cluster to be novel if it contains at least one new or discarded
metric. 
% In particular, the 
% cluster reduction for the top 5 components identified at step 2 is 
% from 25 to 17 (32\%). 
Considering only novel clusters reduces the number of metrics 
and the number of edges for the developers to analyze for the root cause in step 4.
%for RCA include in the final $\{$\textit{component}, \textit{metrics list}$\}$ 
%pairs as well as the number of edges to analyze in step 4. 
%In this case, a cluster is considered to be novel if it contains (at least) one new or discarded 
%metric. 
We also compute the similarity scores for these novel clusters and use the similarity in the next step.

\myparagraph{Step \#4: Edge filtering} By investigating the novel edges
(i.e., new or deleted) in the dependency graph, the developers can better focus on understanding which
component might be more relevant to the root cause. Utilizing different cluster similarity scores
enables developers to filter out some of the edges that may not be relevant.
Figures \ref{fig:rca-evaluation-clusters}(b \& c) 
show the effect of different cluster similarity thresholds for all components 
in Table~\ref{table:rca-eval-individual-metrics}
when filtering edges. 
Without any similarity thresholds, there are 41 edges of interest, 
corresponding to a set of 13 
components, 29 clusters and 221 metrics that might be relevant to the root cause (Figure \ref{fig:rca-evaluation-clusters}(c)). 
% The cluster difference - 23 to 22 - is due to the 
% absence of a causality edge with one of the clusters. 
A higher threshold reduces the number of the 
$\{$\textit{component}, \textit{metrics list}$\}$ pairs: 
filtering out clusters with inter-version similarity 
scores below 0.50, there are 24 edges of interest, corresponding to 
10 components, 16 clusters and 163 metrics. 

Figure~\ref{fig:rca-evaluation-edge-differences} shows the edges between 
the components at the top-5 rows of 
Table~\ref{table:rca-eval-individual-metrics}, with a similarity threshold of
0.50. Note that one component (i.e., Nova scheduler) was removed by the similarity filter.
Another interesting observation is
that one of the new edges includes a Nova API component cluster, in 
which the \textit{nova-instances-in-state-ACTIVE} 
metric is replaced with \textit{nova-instances-in-state-ERROR}. This change relates
directly to the observed anomaly (i.e., error in VM launch). The other end 
of this edge is a cluster in the Neutron component, which aggregates metrics 
related to VM networking, including a metric named \textit{neutron-ports-in-status-DOWN}. 
This observation indicates a causal relationship between the VM failure and a VM networking 
issue, which is the true root cause of the anomaly.

We also note that similarity a high threshold may filter out useful information. For example, 
the Neutron component cluster with the \textit{neutron-ports-in-status-DOWN} 
metric is removed with similarity thresholds above 0.60. 
%Increasing the similarity threshold above 0.75 yields no edges of interest. 
We leave the study of this parameter's sensitivity to future work.

\myparagraph{Step \#5: Final rankings} The rightmost column on 
Table~\ref{table:rca-eval-individual-metrics} shows the final rankings, considering 
edge filtering step with a 0.50 similarity threshold. 
Figure \ref{fig:rca-evaluation-edge-differences} shows 
%The rankings 
%conserves the order of the component novelty scores shown in Table~\ref{table:rca-eval-individual-metrics}, 
a significant reduction in terms of state to analyze (from a total of 
16 components and 508 metrics to 10 and 163, respectively) because of
the exclusion of non-novel clusters. For example, 
for Nova API, the number of metrics reduces from 59 to 20, for Neutron server 
from 42 to 22. Furthermore, our method 
includes the Neutron component as one of the top 5 components, and 
isolates an edge which is directly related with the true root cause of the 
anomaly.

\subsubsection{Bug 1590179 : Fernet token performance regression}
\label{subsubsec:root-cause-evaluation-1533942}

The main symptom of bug \#1590179 is a general decrease in the rate 
at which Openstack processes user requests, in-between 
Openstack `Liberty' and `Mitaka' releases.

\myparagraph{Root cause} As reported in~\cite{launchpad-bug-1590179}, the 
issue is due to a 5$\times$ increase in authentication token validation time. The 
ultimate cause of the bug is a change in the token caching strategy in-between 
Openstack `Liberty' and `Mitaka' releases.

In the context of Openstack, tokens represent the authenticated identity of a specific 
requester (e.g. a system user) and grants authorization for a specific 
Openstack action (e.g. starting a VM)~\cite{openstack-tokens}. Openstack 
supports different types of tokens, but this issue is particular to 
Fernet tokens~\cite{openstack-tokens}, which do not require persistence in a 
database: validation is based on symmetric encryption, with secret keys kept 
by the Openstack identity component. 

\myparagraph{Experimental setup} Similarly to the bug \#1533942 use case,
we deployed OpenStack components as containerized microservices using Kolla~\cite{openstack-kolla}. 
We deployed 7 Openstack components, along with 12 auxiliary components, as listed 
in Table~\ref{table:rca-eval-components}. 
The Openstack versions used for the correct (C) and faulty (F) versions are listed 
in Table~\ref{table:rca-eval-openstack-versions}. The configurations for the Kolla deployments of 
each version are publicly available~\footnote{\href{https://github.com/sieve-microservices/kolla}{https://github.com/sieve-microservices/kolla}}, 
as well as the monitoring infrastructure and evaluation scripts~\footnote{\href{https://github.com/sieve-microservices/rca-evaluation}{https://github.com/sieve-microservices/rca-evaluation}}.

\begin{table}
%\scriptsize
\myfontsize
\centering
\caption{Components deployed by Openstack Kolla during RCA evaluation.}
%\vspace{-2mm}
% make this more complete
\label{table:rca-eval-components}
\newcommand{\tabitem}{~~~~~\llap{\textbullet}~}
\begin{tabular}{@{}cccc@{}}
\toprule
\textbf{Component}    & \textbf{Purpose}  & \textbf{\# Microservices} \\ \midrule
Nova                & VM computing      & 8 \\
Neutron             & VM networking     & 6 \\
Keystone            & Identity          & 3 \\
Glance              & VM image manag.   & 2 \\
Heat                & -                 & 3 \\
Horizon             & Web UI            & 1 \\
Ceilometer          & Telemetry         & 5 \\ \midrule
Heka                & Logging           & 3 \\
Cron                & Job scheduling    & 3 \\
Open vSwitch        & VM networking (aux.) & 4 \\
Elasticsearch       & Search engine     & 1 \\
Kibana              & Data visualiz.    & 1 \\
\multirow{2}{*}{Memcached}  & Auth. token caching       & \multirow{2}{*}{1} \\
                            & (among others)            & \\
\multirow{2}{*}{Mariadb}    & Openstack                 & \multirow{2}{*}{1} \\
                            & parameter storage         & \\
RabbitMQ            & Message broker                    & 1 \\
MongoDB             & Ceilometer data storage           & 1 \\
Telegraf            & Metric collection                 & 1 \\
InfluxDB            & \multirow{2}{*}{Metric storage}   & 1 \\
PostgreSQL          &                                   & 1 \\ \midrule
\textbf{Totals}     & -                                 & \textbf{47} \\  
\bottomrule
\end{tabular}
%\vspace{-3mm}
\end{table}

\begin{table}[t]
%\scriptsize
\myfontsize
\centering
\caption{Details about Rally tasks used in RCA evaluation.}
%\vspace{-2mm}
\label{table:rca-eval-rally-tasks}
\newcommand{\tabitem}{~~~~~\llap{\textbullet}~}
\begin{tabular}{@{}l>{\raggedright}p{2.0cm}cc>{\raggedright}p{1.5cm}c@{}}
\toprule
\textbf{Bug \#} & \textbf{Benchmark}    & \textbf{\# Runs}  & \textbf{Concurr.}  & \textbf{Details} & \\ \midrule

\multirow{2}{*}{1533942}    & boot and delete (\textbf{B\&D}) 
                            & \multirow{2}{*}{100}  & \multirow{2}{*}{5} 
                            & VMs up for 15-25 sec          & \\ \midrule
\multirow{11}{*}{1590179}   & \multirow{2}{*}{\textbf{B\&D}}  
                            & \multirow{2}{*}{25}   & \multirow{2}{*}{5} 
                            & VMs up for 15-25 sec          & \\ \cmidrule[0.5pt]{2-6}
                            & authenticate user and validate token (\textbf{AU\&VT}) 
                            & \multirow{3}{*}{100}  & \multirow{3}{*}{5} 
                            & \multirow{3}{*}{-}            & \\ \cmidrule[0.5pt]{2-6}
                            & create and delete networks (\textbf{C\&DN})               
                            & \multirow{3}{*}{50}   & \multirow{3}{*}{5} 
                            & \multirow{3}{*}{-}            & \\ \cmidrule[0.5pt]{2-6}
                            & create and delete image (\textbf{C\&DI})                  
                            & \multirow{3}{*}{50}   & \multirow{3}{*}{2} 
                            & Cirros 0.35 x86\_64 image     & \\
\bottomrule
\end{tabular}
%\vspace{-3mm}
\end{table}

\begin{table}
%\scriptsize
\myfontsize
\centering
\caption{OpenStack components, sorted by the number of novel metrics 
between {correct (C)} and faulty (F) versions.}
%\vspace{-2mm}
% Final rankings in last 
%column (after edge filtering with similarity threshold of 0.50).}
%Last 8 components include Keystone, Heat, 
%HAProxy as well as subcomponents of Nova, Neutron and Glance.}
\label{table:rca-eval-individual-metrics}
\newcommand{\tabitem}{~~~~~\llap{\textbullet}~}
\begin{tabular}{@{}cccc@{}}
\toprule
\multirow{2}{*}{\textbf{Component}}   & \textbf{Changed}          & \textbf{Total}    & \textbf{Final}    \\ 
                                    & (New\slash Discarded)     & (per component)     & \textbf{ranking}  \\ \midrule
Nova API            & 29 (7\slash 22)       & 59    & 1     \\
Nova libvirt        & 21 (0\slash 21)       & 39    & 2     \\
Nova scheduler      & 14 (7\slash 7)        & 30    & -     \\
Neutron server      & 12 (2\slash 10)       & 42    & 3     \\
RabbitMQ            & 11 (5\slash 6)        & 57    & 4     \\
Neutron L3 agent    & 7 (0\slash 7)         & 39    & 5     \\
Nova novncproxy     & 7 (0\slash 7)         & 12    & -     \\
Glance API          & 5 (0\slash 5)         & 27    & 6     \\
Neutron DHCP ag.    & 4 (0\slash 4)         & 35    & 7     \\
Nova compute        & 3 (0\slash 3)         & 41    & 8     \\
Glance registry     & 3 (0\slash 3)         & 23    & 9     \\
Haproxy             & 2 (1\slash 1)         & 14    & 10    \\
Nova conductor      & 2 (0\slash 2)         & 29    & -     \\
Other 3 components    & 0 (0\slash 0)         & 59    & -     \\ \midrule
\textbf{Totals}     & \textbf{113} (\textbf{22}\slash \textbf{91})   & \textbf{508} & - \\  
\bottomrule
\end{tabular}
%\vspace{-3mm}
\end{table}

The infrastructure consists of 3 t2.large Amazon EC2 VM instances to run OpenStack components (2 vCPUs, 8 GB RAM and 
30 GB Amazon EBS storage) and 2 t2.medium VM instances (2 vCPUs, 4GB RAM and 
30 GB EBS storage) for the supporting components (metric collection and storage).

\myparagraph{Results} Since this bug manifests itself as a general performance degradation issue, we 
run the Rally tasks below to load 4 essential Openstack components, thus 
giving us a reasonably large `search space' for RCA (details in Table~\ref{table:rca-eval-rally-tasks}):

\begin{itemize}
\item \textbf{B\&D}: Same Rally task used in bug \#1533942. Loads 
Openstack's compute (Nova) and networking (Neutron) components.
\item \textbf{AU\&VT}: Authenticates a stream of 
user's tokens in Keystone, Openstack's identity component. In hindsight, since the reported root cause of 
bug \#1590179 is related to Keystone~\cite{launchpad-bug-1590179}, this might appear as a `dishonest' test. However, we argue 
that Keystone is a central component of Openstack - as are the compute and networking components - and 
as such a natural candidate for testing.
\item \textbf{C\&DN}: Creates and deletes network VM network resources. Loads 
Openstack's networking component - Neutron - as well as related components (e.g. Open vSwitch).
\item \textbf{C\&DI}: Creates and deletes VM images. Loads Openstack's image
component, Glance.
\end{itemize}

\begin{table}
%\scriptsize
\myfontsize
\centering
\caption{Openstack versions used in RCA evaluation.}
%\vspace{-2mm}
\label{table:rca-eval-openstack-versions}
\newcommand{\tabitem}{~~~~~\llap{\textbullet}~}
\begin{tabular}{@{}lcll@{}}
\toprule
\textbf{Bug \#} & \textbf{Correctness}    & \textbf{Openstack ver.}  & \textbf{Comments} \\ \midrule

\multirow{6}{*}{1533942}    & \multirow{4}{*}{Correct (C)}  & \multirow{3}{*}{Mitaka (EOL)}     & Adapted for Ceilometer \\
                            &                               &                                   & support (not available by \\ 
                            &                               &                                   & default)\textsuperscript{1} \\
                            &                               & Kolla 2.0.0.0b3                   & - \\ \cmidrule[0.5pt]{2-4}
                            & \multirow{2}{*}{Faulty (F)}   & Mitaka (EOL)                      & (1) \\
                            &                               & Kolla 2.0.0.0b2                   & - \\ \midrule
\multirow{5}{*}{1590179}    & \multirow{3}{*}{Correct (C)}  & \multirow{1}{*}{Liberty (EOL)}    & (1) \\
                            &                               & \multirow{2}{*}{Keystone 8.1.0}   & Adapted to Fernet tokens \\
                            &                               &                                   & based on blueprint in~\cite{openstack-kolla-fernet-blueprint}\textsuperscript{2} \\ \cmidrule[0.5pt]{2-4}
                            & \multirow{2}{*}{Faulty (F)}       & Mitaka (EOL)      & (1) \\
                            &                                   & Keystone 9.0.2    & (2) \\
\bottomrule
\end{tabular}
%\vspace{-3mm}
\end{table}

\begin{table*}
%\scriptsize
\myfontsize
\centering
\caption{Bug \#151590179 results: OpenStack components, sorted by the number of novel metrics 
between {correct (C)} and faulty (F) versions, for different Rally tasks. Metric novelty rankings correspond to column `N', 
final rankings (after edge filtering steps) correspond to column `F'. Components which are ranked 1st according to the `metric (N)ovelty' ranking are \underline{underlined}, those ranked 1st after edge filtering (`(F)inal' rankings) are emphasized in \textbf{bold}.}
%\vspace{-2mm}
% Final rankings in last 
%column (after edge filtering with similarity threshold of 0.50).}
%Last 8 components include Keystone, Heat, 
%HAProxy as well as subcomponents of Nova, Neutron and Glance.}
\label{table:rca-eval-individual-metrics-bug-1590179}
\newcommand{\tabitem}{~~~~~\llap{\textbullet}~}
\begin{tabular}{@{}c ccc c ccc c ccc c ccc cccc@{}}
\toprule
\multirow{4}{*}{\textbf{Component}}   & \multicolumn{19}{c}{\textbf{Rally tasks}} \\ \cmidrule[0.5pt]{2-20}
                                    & \multicolumn{4}{c}{AU\&VT} & 
                                    & \multicolumn{4}{c}{B\&D} & 
                                    & \multicolumn{4}{c}{C\&DN} & 
                                    & \multicolumn{4}{c}{C\&DI} \\ \cmidrule[0.5pt]{2-5}\cmidrule[0.5pt]{7-10}\cmidrule[0.5pt]{12-15}\cmidrule[0.5pt]{17-20}
                                    & (C)hanged             & (T)otal               & \multicolumn{2}{c}{Rankings} & 
                                    & \multirow{2}{*}{C}    & \multirow{2}{*}{T}    & \multirow{2}{*}{N} & \multirow{2}{*}{F}  & 
                                    & \multirow{2}{*}{C}    & \multirow{2}{*}{T}    & \multirow{2}{*}{N} & \multirow{2}{*}{F} & 
                                    & \multirow{2}{*}{C}    & \multirow{2}{*}{T}    & \multirow{2}{*}{N} & \multirow{2}{*}{F} \\ 
                                    & (New\slash Disc.) & & N & F & & & & & & & & & & & & & & & \\ \midrule
\underline{\textbf{Neutron OvSwitch}} & \underline{42 (15\slash 27)} & \underline{53} & \underline{1} & \underline{-} & 
                                    & 0 (0\slash 0)             & 38            & 14 & -                  & 
                                    & \underline{\textbf{40 (14\slash 26)}} & \underline{\textbf{52}} & \underline{\textbf{1}} & \underline{\textbf{1}} & 
                                    & \underline{42 (15\slash 27)} & \underline{53} & \underline{1} & \underline{-}   \\
\textbf{Memcached}                  & \textbf{15 (14\slash 1)}  & \textbf{30}   & \textbf{2} & \textbf{1}      & 
                                    & 6 (4\slash 2)             & 33            & 5 & 3               & 
                                    & 7 (3\slash 3)             & 33            & 10 & 4              & 
                                    & 5 (4\slash 1)             & 31            & 11 & 3              \\
\textbf{Nova API}                   & 10 (10\slash 0)           & 25            & 3 & -               & 
                                    & 3 (3\slash 0)             & 44            & 10 & 6              & 
                                    & 20 (15\slash 5)           & 32            & 3 & -               & 
                                    & \textbf{18 (13\slash 5)}  & \textbf{30}   & \textbf{3} & \textbf{1}      \\
Nova conductor                      & 9 (0\slash 9)             & 15            & 4 & 2               & 
                                    & 2 (2\slash 0)             & 30            & 12 & 7              & 
                                    & 27 (0\slash 27)           & 33            & 2 & -               & 
                                    & 0 (0\slash 0)             & 6             & 18 & -              \\
Nova libvirt                        & 8 (0\slash 8)             & 17            & 5 & 3               & 
                                    & 2 (2\slash 0)             & 39            & 13 & 8              & 
                                    & 7 (1\slash 6)             & 17            & 8 & -               & 
                                    & 12 (6\slash 6)            & 23            & 5 & 2               \\
\textbf{Neutron server}             & 6 (6\slash 0)             & 21            & 5 & -               & 
                                    & \textbf{12 (5\slash 7)}   & \textbf{52}   & \textbf{2} & \textbf{1}      & 
                                    & 6 (4\slash 2)             & 34            & 11 & 5              & 
                                    & 18 (13\slash 5)           & 30            & 4 & -               \\
Neutron L3 agent                    & 4 (0\slash 4)             & 19            & 7 & -               & 
                                    & 6 (4\slash 2)             & 46            & 4 & -               & 
                                    & 0 (0\slash 0)             & 15            & 16 & -              & 
                                    & 2 (2\slash 0)             & 15            & 13 & -              \\
Nova SSH                            & 3 (3\slash 0)             & 3             & 8 & -               & 
                                    & - (-\slash -)             & -             & - & -               & 
                                    & 3 (3\slash 0)             & 3             & 14 & -              & 
                                    & - (-\slash -)             & -             & - & -               \\
Glance registry                     & 3 (1\slash 2)             & 45            & 9 & -               & 
                                    & 0 (0\slash 0)             & 30            & 17 & -              & 
                                    & 7 (7\slash 0)             & 7             & 9 & -               & 
                                    & 8 (8\slash 0)             & 8             & 7 & -               \\
RabbitMQ                            & 3 (3\slash 0)             & 3             & 10 & 4              & 
                                    & 4 (2\slash 2)             & 63            & 6 & 4               & 
                                    & 8 (8\slash 0)             & 54            & 6 & 2               & 
                                    & 10 (9\slash 1)            & 57            & 6 & 3               \\
Neutron DHCP ag.                    & 2 (2\slash 0)             & 15            & 11 & -              & 
                                    & 0 (0\slash 0)             & 42            & 15 & 9              & 
                                    & 0 (0\slash 0)             & 15            & 18 & -              & 
                                    & 0 (0\slash 0)             & 15            & 17 & -              \\
Nova compute                        & 2 (0\slash 2)             & 19            & 12 & -              & 
                                    & 0 (0\slash 0)             & 45            & 16 & -              & 
                                    & 6 (2\slash 4)             & 23            & 12 & -              & 
                                    & 2 (0\slash 2)             & 19            & 14 & -              \\
Nova novncproxy                     & 1 (1\slash 0)             & 4             & 13 & -              & 
                                    & 3 (0\slash 3)             & 17            & 11 & -              & 
                                    & 8 (7\slash 1)             & 12            & 7 & 3               & 
                                    & 8 (8\slash 0)             & 13            & 8 & -               \\
Nova scheduler                      & 0 (0\slash 0)             & 6             & 14 & -              & 
                                    & 4 (1\slash 3)             & 31            & 8 & -               & 
                                    & 14 (14\slash 0)           & 14            & 5 & -               & 
                                    & 0 (0\slash 0)             & 14            & 15 & -              \\
\underline{Nova consoleauth}        & 0 (0\slash 0)             & 6             & 15 & -              & 
                                    & \underline{18 (4\slash 14)} & \underline{36} & \underline{1} & \underline{-} & 
                                    & 0 (0\slash 0)             & 6             & 17 & -              & 
                                    & 5 (5\slash 0)             & 20            & 10 & -              \\
Keystone                            & 0 (0\slash 0)             & 36            & 16 & -              & 
                                    & 8 (6\slash 2)             & 42            & 3 & 2               & 
                                    & 2 (1\slash 1)             & 37            & 15 & 7              & 
                                    & 0 (0\slash 0)             & 36            & 16 & -              \\
Neutron metadata ag.                & 0 (0\slash 0)             & 16            & 17 & -              & 
                                    & 0 (0\slash 0)             & 24            & 18 & -              & 
                                    & 16 (16\slash 0)           & 16            & 4 & -               & 
                                    & 20 (20\slash 0)           & 20            & 2 & -               \\
Glance API                          & - (-\slash -)             & -             & 18 & -              & 
                                    & 4 (0\slash 4)             & 34            & 7 & 5               & 
                                    & 5 (5\slash 0)             & 15            & 13 & 6              & 
                                    & 6 (6\slash 0)             & 6             & 9 & -               \\ \midrule
\textbf{Totals}                     & \textbf{103} (\textbf{50}\slash \textbf{53})  & \textbf{332} & - & - & 
                                    & \textbf{76}  (\textbf{34}\slash \textbf{42})  & \textbf{657} & - & - &
                                    & \textbf{136} (\textbf{61}\slash \textbf{75})  & \textbf{378} & - & - &
                                    & \textbf{128} (\textbf{81}\slash \textbf{47})  & \textbf{392} & - & - \\
\bottomrule
\end{tabular}
%\vspace{-3mm}
\end{table*}

% \begin{figure*}
% \begin{tabular}{c}
% \subfloat[{AU\&VT Rally task results: {\bf (a)} Cluster novelty score. {\bf (b)} Edge novelty score.  {\bf (c)} No. of components \& clusters after edge filtering w/ varying thresholds.}]{\includegraphics[width=0.95\textwidth]{figures/rca-evaluation-clusters-bug-1590179-auvt.pdf}}
% % \\
% % \subfloat[{B\&D Rally task results}]{\includegraphics[width=0.95\textwidth]{figures/rca-evaluation-clusters-bug-1590179-bd.pdf}}\\
% % \subfloat[{C\&DN Rally task results}]{\includegraphics[width=0.95\textwidth]{figures/rca-evaluation-clusters-bug-1590179-cdn.pdf}}\\
% % \subfloat[{C\&DI Rally task results}]{\includegraphics[width=0.95\textwidth]{figures/rca-evaluation-clusters-bug-1590179-cdi.pdf}}
% % \subfloat[{\scriptsize Events \& outcome prob. for max. aggregation and different req. sizes (256 bit BFs, `flooding').}]{\includegraphics[width=0.30\textwidth]{figures/req-sizes.pdf}}
% \end{tabular}
%     \caption{RCA evaluation results for bug \#151590179, using different Rally tasks.}
%     \label{fig:rca-evaluation-edge-differences-bug-151590179}
% \end{figure*}

\begin{figure*}[t]
\centering
\includegraphics[scale=.425]{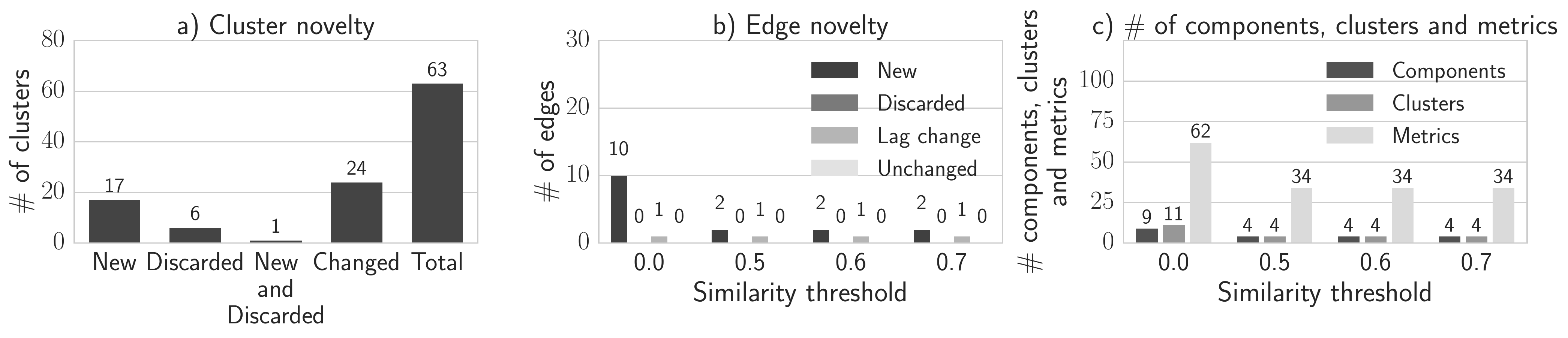}
\caption{AU\&VT Rally task results: {\bf (a)} cluster novelty score. {\bf (b)} edge novelty score.  {\bf (c)} no. of components \& clusters after edge filtering w/ varying thresholds.}
\label{fig:rca-evaluation-edge-differences-bug-151590179}
\vspace*{-0.4cm}
\end{figure*}

\myparagraph{Metric analysis and component rankings} 
Table \ref{table:rca-eval-individual-metrics-bug-1590179} shows that 
solely based on metric novelty scores, the `authenticate user and validade token' (AU\&VT) 
Rally task - which directly relates to token validation - ranks Memcached 2nd place
as a possible root cause (metric novelty rankings correspond to column `N' in 
Table~\ref{table:rca-eval-individual-metrics-bug-1590179}, Memcached is ranked 1st after edge filtering, i.e. the 
`F' column). 
Other tasks rank Nova conosoleauth (B\&D) and Neutron OpenvSwitch agent (C\&DN and C\&DI) as 1st, which 
are unrelated to bug \#1590179 (at least not directly related to it, according 
to the description given in~\cite{launchpad-bug-1590179}).

Still regarding metric novelty, Keystone - one of the Openstack components directly related to 
bug \#1590179~\cite{launchpad-bug-1590179} - is lowly ranked according to the 
AU\&VT Rally task. This result is also intuitive: since bug \#1590179 
relates to performance degradation, metrics should not appear or 
disappear, rather their values should differ in-between C and F versions, 
an effect which may be captured by relationships in-between the clusters 
the metrics belong to.

\begin{table}
%\scriptsize
\myfontsize
\centering
\caption{Change in `metric novelty' rankings for bugs \#1533942 and 
\#1590179, considering the `boot and delete' Rally task.}
%\vspace{-2mm}
% Final rankings in last 
%column (after edge filtering with similarity threshold of 0.50).}
%Last 8 components include Keystone, Heat, 
%HAProxy as well as subcomponents of Nova, Neutron and Glance.}
\label{table:rca-eval-rank-change-bug-1590179}
\newcommand{\tabitem}{~~~~~\llap{\textbullet}~}
\begin{tabular}{@{}cccc@{}}
\toprule
\multirow{2}{*}{\textbf{Component}}     & \multicolumn{2}{c}{\textbf{Rank (bug \#})}    & \multirow{2}{*}{\textbf{Ranking change}} \\ 
                                        & 1533942 & 1590179                             & \\ \midrule
Nova API            & 1     & 10        & -9        \\
Nova libvirt        & 2     & 11        & -11       \\
Nova scheduler      & 3     & 8         & -5        \\
Neutron server      & 4     & 2         & +2        \\
RabbitMQ            & 5     & 6         & -1        \\
Neutron L3 agent    & 6     & 4         & +2        \\
Nova novncproxy     & 7     & 11        & -4        \\
Glance API          & 8     & 7         & +1        \\
Neutron DHCP ag.    & 9     & 15        & -6        \\
Nova compute        & 10    & 16        & -6        \\
Glance registry     & 11    & 17        & -6        \\
Nova conductor      & 13    & 12        & +1        \\
Keystone            & 15    & 3         & +12       \\ \bottomrule
\end{tabular}
%\vspace{-3mm}
\end{table}

Since we load Openstack with the `boot and delete' (B\&D) tasks in both 
bugs \#1533942 and \#1590179, we can compare how the rankings change between 
bugs, and verify if there is evidence of a a dependency between the way 
in which the system is loaded and the `metric novelty' rankings produced 
by \sys. As shown in Table~\ref{table:rca-eval-rank-change-bug-1590179}, the average 
change in rank is $\sim5$ positions, providing initial evidence about 
the lack of such dependency. Further validation of 
this dependency is left to future work.

% \myparagraph{Step \#3: Cluster novelty \& similarity} 
\myparagraph{Edge filtering} The edge novelty statistics results for the 
AU\&VT task - depicted in Figures~\ref{fig:rca-evaluation-edge-differences-bug-151590179}(b) - are of 
particular interest, since it directly loads Openstack with a task related to bug \#1590179.

The edge filtering step identifies 10 new edges no inter-version cluster similarity threshold 
is applied. These reduce to 2 once the similarity threshold is raised to 0.5. The 2 new 
edges are between the following components: RabbitMQ > Nova conductor and Memcached > Nova 
libvirt. The Memcached metrics associated with the edge are not related to 
cache access metrics, e.g. cache hits or misses. The single edge 
isolated due to a causality `lag change' pertains to a relationship between 
RabbitMQ and and Nova libvirt, with no apparent relation to the bug. 
None of the filtered relationships involves Keystone, and as such metrics such as 
`keystone identity authenticate success rate', which are known to be related to 
the bug.

\myparagraph{Final rankings} In the case of the AU\&VT task, Memcached - which effectively is related to the bug - is ranked 1st after the 
edge filtering steps. However, none of the metrics filtered in the edge filtering steps 
seems to directly relate to bug \#1590179. Also, an intuitive isolation of lag changes in 
edges involving Keystone clusters did not occur. Further experimentation is required to 
assess \sys's effectiveness for RCA of performance degradation issues (as opposed to 
crashes, such as bug \#1533942) such bug \#1590179.

%% file: related-work.tex
% I think I forgot to mention one paper you used to list here; how can I find a previous version of this tex file?

\section{Related Work}

\myparagraph{Scalable Monitoring} With the increasing number of metrics 
exposed by distributed cloud systems, the scalability of the monitoring process 
becomes crucial. Meng et al.~\cite{meng2013tc} optimize monitoring scalability by 
choosing appropriate monitoring window lengths and adjusting the monitoring 
intensity at runtime. Canali et al.~\cite{canali2014adaptive} achieve scalability 
by clustering metric data. A fuzzy logic approach is used to speed up 
clustering, and thus obtain data for decision making within shorter periods.
Rodrigues et al.~\cite{da2015interplay} explore the trade-off between 
timeliness and the scalability in cloud monitoring, and analyze the mutual 
influence between these two aspects based on the monitoring parameters. Our work is complementary to existing monitoring systems since \sys aims to improve the efficiency by monitoring less number of metrics. 
%, such as monitoring topologies and frequency.

\myparagraph{Distributed Debugging}
Systems like Dapper \cite{dapper} and Pip \cite{pip}
require the developers to instrument the application to obtain its causal model.
X-trace \cite{x-trace} uses a modified network stack to propagate useful information
about the application. In contrast, \sys does not modify the application code to
obtain the call/dependency graph of the application.

Systems such as Fay \cite{erlingsson2011} and DTrace \cite{dtrace} 
enable developers to dynamically inject debugging
requests by developers and require no initial logs of metrics. Pivot Tracing \cite{mace15}
combines dynamic instrumentation with causality tracing.
\sys can complement these approaches, because it can
provide information about interesting components and metrics, so that the developers
can focus their efforts to understand them better. 
Furthermore, \sys's dependency graph
is a general tool that can not only be used for debugging, but also for other purposes
such as orchestration~\cite{conductor-nsdi-2012, conductor-ladis-2010, conductor-podc-2010}.

Data provenance~\cite{lpm, inspector, spade} is another technique that can be used to trace the dataflow in the system. \sys can also leverage the existing provenance tools to derive the dependence graph.

\myparagraph{Metric reduction} Reducing the size and dimensionality of the bulk 
of metric data exposed by complex distributed systems is essential for its 
understanding. Common techniques include sampling, and data clustering via $k$-means and $k$-medoids.
Kollios et al.~\cite{kollios2003efficient} employ biased sampling to capture 
the local density of datasets. Sampling based approaches argues for approximate computing~\cite{incapprox, privapprox, streamapprox-middleware} to enable a systematic trade-off between the accuracy, and efficiency to collect and compute on the metrics. Zhou et al.~\cite{zhou2000combining}
simply use random sampling due to its simplicity and low complexity. 
% For data clustering, $k$-means and $k$-medoids are typical partitioning algorithms
% which require the prior specification of $k$. For instance, 
Ng et al.~\cite{vldb94}
improved the $k$-medoid method and made it more effective and efficient.
Ding et al.~\cite{ding2015yading} rely on clustering over sampled 
data to reduce clustering time. 

% The non-sampled data is assigned to the appropriate 
%clusters, with guarantees of distribution consistency between the input and 
%and sampled data.

%[VLDB'15] YADING-Fast-Clustering-of-Large-Scale-Time-Series-Data
%[VLDB'94] Efficient and Effective Clustering Methods for Spatial Data Mining
%[TKDE'03] Efficient biased sampling for approximate clustering and outlier detection in large datasets
%[PAKDD'00] Combining sampling technique with DBSCAN algorithm for clustering large spatial databases

\sys's approach is unique because of its two-step approach: (1) we first cluster 
time series to identify the internal dependency between \emph{any} 
given metrics
% of a distributed system; 
and then (2) infer the causal 
relations among time series. Essentially, \sys uses two steps of data reduction
for better reduction.
% offering better performance in terms of data reduction. 
Furthermore, \sys's time series 
processing method extracts other useful information such as the time delay of 
the causal relationship, which can be leveraged in different use cases (e.g., root cause analysis).
% when applying our approach to difference cases.

%[SIGCOMM'16] The Good, the Bad, and the Differences: Better Network Diagnostics with Differential Provenance
%[CoNEXT'15] Hansel: Diagnosing Faults in Openstack
%[SIGMETRICS'13] Root Cause Detection in a Service-Oriented Architecture
%[CNSM'2010] Dependency-aware fault diagnosis with metric-correlation models in enterprise software systems
% THIS IS TOO LONG...

\myparagraph{Orchestration of autoscaling} Current techniques for autoscaling can be broadly classified into four 
categories~\cite{botran2014}: 
{\em (i)} {static and threshold-based rules} (offered by most cloud computing providers~\cite{aws,heat,azure,gcloud});
{\em (ii)}   {queuing theory} \cite{ali-eldin2012, zhang2007, han2014}; 
{\em (iii)}  {reinforcement learning}~\cite{yazdanovcloud2014, tesauro2006, rao2009}; and 
{\em (iv)}  {time series analysis}~\cite{khatua2010, chen2008, roy2011}.
%{\em (iv)}  {time series analysis}~\cite{khatua2010, chen2008, roy2011, huang2012, caron2010, mi2010}.
%Threshold-based rules put upper and lower limits on a set of selected metrics to
%decide when to scale up or down. It is used by most cloud computing solutions, e.g. 
%Amazon AWS~\cite{aws}, OpenStack Heat~\cite{heat}, Windows Azure~\cite{azure}
%and Google Cloud~\cite{gcloud}. Despite its simplicity, it requires
%application knowledge to pick appropriate metrics and thresholds. 
%
%Queuing theory uses a mathematical models of queues. It allows 
%for the prediction of the mean response time of a system, given the knowledge 
%of certain parameters (e.g. request arrival rate, service rate) \cite{ali-eldin2012, zhang2007, han2014} . 
%% To model microservices with multiple applications a queuing network can be constructed.
%One challenge with using queuing theory for microservice autoscaling 
%is that the queuing model has to be frequently updated, given the volatility 
%of load. 
%
%Reinforcement learning is an unsupervised learning method, which tries to find the best
%scaling decision for every system state by learning from past interactions~\cite{yazdanovcloud2014, tesauro2006, rao2009}.
%There are also proposal that uses time series analysis based on past measurements, represented as data points indexed
%in time order, to find repeating patterns in the past or forecast future values~\cite{khatua2010, chen2008, roy2011, huang2012, caron2010, mi2010}.
%It can be used to predict future workload and plan rules to proactively scale
%applications. 
Existing systems using these techniques
can benefit from the selection of better metrics and/or from the dependencies 
between components. In this regard, our work is complementary to these techniques: it is intended to provide the developers 
with knowledge about the application as a whole. In our case study, we showed the benefits of \sys 
for an autoscaling engine using threshold-based rules.

\myparagraph{Root Cause Analysis (RCA)} Large and complex distributed systems are susceptible to anomalies, whose
root causes are often hard to diagnose~\cite{gremlin}.
% Metric correlation for RCA has been proposed in the past.
Jiang et al.\ \cite{rca-2010} compare ``healthy" and ``faulty" metric 
correlation maps, searching broken correlations. In contrast, \sys leverages Granger causality 
instead of simple correlation, allowing for richer causality inference (e.g., causality direction,
time lag between metrics).
MonitorRank~\cite{monitorrank} uses metric collection for RCA in a service-oriented 
architecture.
%, assuming a previous anomaly detection. 
It only analyzes pre-established (component, metric) relations according to a previously-generated
call graph.
%The analyzed metric relations 
%are restricted according to a previously-generated call graph (i.e., pre-established (service, metric)
%relations). 
\sys also uses a call graph, but does not fix metric relations between components, 
for a richer set of potential root causes.
%DIFFPROV~\cite{diffprov} identifies root causes by looking at the differences
%between provenance graphs. \sys's approach is similar, looking at differences between 
%dependency graphs. Unlike \sys, DIFFPROV is supported by a
%provenance system, i.e. instrumentation which tracks the causal connections 
%between states and events. We build dependency graphs based on 
%system metrics, exposed by default, without the need for instrumentation.
There are other application-specific solutions for RCA (e.g., Hansel \cite{openstack-hansel}, Gretel \cite{openstack-gretel}). In contrast, \sys uses a general approach for understanding 
the complexity of microservices-based applications that can support RCA as well as other use cases.

%% file: lessons.tex
\section{Experience and Lessons Learned}
\label{sec:lessons}

While developing \sys, we set ourselves ambitious design goals (described in $\S$\ref{subsec:goals}). However, we learned the following lessons while designing and deploying \sys for real-world applications. 

\myparagraph{Lesson\#1} 
When we first designed \sys, we were envisioning a dependency graph that was clearly showing the relationships between components (e.g., a tree). As a result,
not only would the number of metrics that needed to be monitored be reduced, but also the number of components: one would only need to observe the root(s) of the dependency graph, and make the actions of the dependent components according to the established relationships between the root(s) and them. Such a dependency graph would give the orchestration scenario a huge benefit. Unfortunately, our experience has shown us that the relationships between components are usually not linear, making the dependency graph more complex. Also, there was no obvious root. Consequently, we had to adjust our thinking and utilize some application knowledge regarding components and their relations with others.
% Although, we would like to have one panacea for all diseases---it turned out difficult for us to fully automate the solution without domain knowledge. In particular, based on \sys's output, we would like to automate the development of management tools for a wide-range of workflows targeting efficiency, resiliency, and dependability in distributed systems. However, our platform still requires specific knowledge from the application domain to effectively utilize \sys's output. For instance, the root cause analysis scenario required the knowledge of a faulty and non-faulty version of the application. 
Nevertheless, in our experience, \sys provides the developer with a good starting point to improve their workflows. 

\myparagraph{Lesson\#2} \sys is designed for ``blackbox'' monitoring of the evaluated application, where \sys can collect and analyze generic system metrics that are exposed by the infrastructure (e.g., CPU usage, disk I/O, network bandwidth).  However, in our experience, a system for monitoring and analyzing an application should also consider application-specific metrics  (e.g., request latency, number of error messages) to build effective management tools. Fortunately, many microservices applications we analyzed already export such metrics. However, given the number of components and exported metrics, this fact can easily create an ``information overload'' for the application developers. In fact, the main motivation of \sys was to deal with this ``information overload''.  Our experience showed that \sys can still monitor the application in the blackbox mode (i.e., no instrumentation to the application), but also overcome the barrage of application-specific metrics.

\myparagraph{Lesson\#3} To adapt to the application workload variations, \sys needs to build a robust model for the evaluated application. This requires a workload generator that can stress-test the application thoroughly. To meet this requirement, there are three approaches: (1) In many cases the developers already supply an application-specific workload generator. For instance, we employed the workload generator shipped with the OpenStack distribution. (2) For cases where we did not have an existing workload generator, we implemented a custom workload generator for the evaluated application. For example, we built a workload generator for ShareLatex. Although we were able to faithfully simulate user actions in ShareLatex, such an approach might not be feasible for some applications. Having the ability to utilize existing production traces (e.g., by replaying the trace or by reproducing similar traces) or working in an online fashion to generate the model of the application would certainly help \sys. Custom workload generation can then be used to close the gaps in the model for certain workload conditions not covered by the existing traces.
% \iea{where did this come from? i don't remember this. we can probably remove it.}
(3) We could also explore some principled approaches for automatic workload generation, such as symbolic execution in distributed systems \cite{achilies}. 
%For example, 
%symbolic execution in distributed systems~\cite{achilies} can be used.
%In particular, one could use symbolic execution in distributed systems~\cite{achilies} to generate workload. 

% \myparagraph{Lesson\#4} For the transparency point of view, 
% {\bf The following assumption is a non-issue to me -- in the cloud environment everyone gets its own VM -- whats the problem?}
% 
% 
% we he following assumptions. First, we assume that the developers have 
% access to the underlying infrastructure, so that they can install and run software on the servers running the
% components of the applications. In microservices-based applications following a DevOps model, 
% this assumption is reasonable because the developers and the operators of the applications are the same or belong to the same organization.

%% file: conclusion.tex
\section{Conclusion and Future Work} This paper reports on our experiences with designing and building \sys, a platform to automatically derive actionable insights from monitored metrics in distributed systems. \sys achieves this goal by automatically reducing the amount of metrics and inferring inter-component dependencies. Our general approach is independent of the application, and can be deployed in an unsupervised mode without prior
knowledge of the time series of metrics. 
We showed that \sys's resulting model is consistent, and can be applied for common use cases such as autoscaling and root-cause debugging.

An interesting research challenge for the future would be to integrate \sys{} into the continuous integration pipeline of an application development. In this scenario, the dependency graph can be updated incrementally~\cite{ithreads,bhatotia15,incoop}, which would speed up the analytics part. In this way, the developers would be able to get real-time profile updates of their infrastructure.  Another challenge is to utilize already existing traffic to generate the dependency graph without requiring the developers to load the system. Using existing traffic would alleviate the burden of developers to supply a workload generator. On the other hand, existing traffic traces might not always capture the stress points of the application. A hybrid approach, in which workload generation is only used for these corner cases, might help to overcome this problem. 

\myparagraph{Software availability} The source code of \sys is publicly available: \href{https://sieve-microservices.github.io/}{https://sieve-microservices.github.io/}.

\myparagraph{Acknowledgments}
We would like to thank Amazon AWS for providing the required infrastructure to run the experiments.